\documentclass {biophys}
\usepackage{helvet,times}
\usepackage{bm,textcomp}
\usepackage[version=3]{mhchem}
\usepackage{graphicx}
\usepackage{subcaption}

\jno{kxl014} %journal number
\gridframe{N}%option for grid around the text "Y" or "N"
\cropmark{N}%option for cropmark around the text "Y" or "N"

\doi{doi: 10.1529/biophysj.106.090944}% DOI number in the copyright line
\usepackage{xr}
\usepackage{amsmath}
\usepackage[round,numbers,sort&compress]{natbib}

\usepackage{units}
\begin{document}

\setcounter{page}{1} %first page number

\title{High-Resolution Structure and Intermolecular Interactions between L-type Straight Flagellar Filaments}

\author{D. Louzon, %$^{\ast,\dagger}$
A. Ginsburg, %$^{\ast,\ddagger}$
W. Schwenger, %$^\S$ 
T. Dvir, %$^{\ast,\dagger}$
Z. Dogic, %$^\S$
and U. Raviv %$^\ast$
}

%\address{$^\ast$ The Institute of Chemistry and the Center for Nanoscience and Nanotechnology, The Hebrew University of Jerusalem, Jerusalem, 91904, Israel.
%$^\dagger$The Racah Institute of Physics, The Hebrew University of Jerusalem, Jerusalem, 91904, Israel.
%$^\ddagger$The Institute for drug research, The Hebrew University of Jerusalem.  
%$^\S$ Department of Physics, Brandeis University, Waltham, MA 02454, USA.}

% generate the title page from the info in the headers above

%Abstract environment needs 3 arguments. They are
%1. The abstract
%2. Received date
%3. Address, email

\begin{abstract}%
{Bacterial mobility is powered by rotation of helical flagellar filaments driven by rotary motors. Flagellin isolated from {\it Salmonella Typhimurium} SJW1660 strain, which differs by a point mutation from the wild-type strain, assembles into straight filaments in which flagellin monomers are arranged into left-handed helix. Using small-angle X-ray scattering (SAXS) and osmotic stress methods, we investigated the high-resolution structure of SJW1660 flagellar filaments as well as intermolecular forces that govern their assembly into dense hexagonal bundle. The scattering data were fitted to high-resolution models, which took into account the atomic structure of the flagellin subunits. The analysis revealed the exact helical arrangement and the super-helical twist of the flagellin subunits within the filaments. Under osmotic stress the filaments formed $2D$ hexagonal bundles. Monte-Carlo simulations and continuum theories were used to analyze the scattering data from hexagonal arrays, revealing how bulk modulus, as well as how the deflection length depends on the applied osmotic stress. Scattering data from aligned flagellar bundles confirmed the predicated structure-factor scattering peak line-shape. Quantitative analysis of the measured equation of state of the bundles revealed the contributions of the electrostatic, hydration, and elastic interactions to the intermolecular interactions associated with bundling of straight semi-flexible flagellar filaments.}%1
{Insert Received for publication Date and in final form Date.}%2
{Corresponding author: % uri.raviv@mail.huji.ac.il
}%3 
\end{abstract}

\maketitle %%The above information typeset through this command

\section*{INTRODUCTION}

Bacterial locomotion is powered by rotating long (ca. $10 - 15\: \mu\textrm{m}$), helical flagellar filaments, which are attached to the bacterial surface  through a molecular motor embedded into the bacterial membrane. The complete flagellum-motor complex contains about $25$ proteins. The flagellin homopolymer, however, comprises more than $99\%$ of the flagellum length providing the structural stiffness necessary to generate thrust that powers bacterial motility~\cite{berg2000,Calladine1975}. Each flagellar filament can be described as a helical assembly of flagellin protein monomers, with nearly 11 subunits per two turns of a 1-start helix, or as a hollow cylinder, comprising 11 protofilaments staggered in a nearly longitudinal helical arrangement~\cite{Asakura1970,Calladine1975}. Each protofilament is a linear structure consisting of flagellin monomers stacked onto each other. 

The flagellin monomers can exist in two distinct conformational states denoted as left (L)- or right (R)-handed. Within each protofilament all the monomers switch in a highly cooperative fashion and thus each protofilament also has either an L or R configuration. If all the protofilaments within a single flagellum have the same conformational state, the entire assemblages assumes a shape of a straight hollow cylinder made of left- or right-handed helical arrangement of the flagellin monomers~\cite{Hasegawa1998}. In many cases, however, a flagellum contains a mixture of L and R protofilaments, leading to a packing frustration that is resolved by the formation of a helical super-structure along the entire flagellum length, a unique feature that is essential for bacterial motility. Depending on the ratio of R and L filaments there are a number of distinct structures of varying pitch and radius. In addition, point mutations in the flagellin amino-acid sequence affect the helical structure~\cite{Hyman1991}. Flagellin mutants, in which all the protofilaments assume L or R conformational state, have been isolated and were shown to assemble into straight flagella filaments~\cite{Kamiya1979,Trachtenberg1991}. Another unique feature of flagella is that it can switch between different helical states in response to external stimuli, including ionic strength, pH, external forces, or temperature~\cite{Asakura1970,Darnton2007,Srigiriraju2005}. Besides their obvious biological importance the unique helical structure and intriguing stimuli induced polymorphic transitions make flagella a highly promising yet poorly explored building block for assembly of soft materials and biologically inspired nano/micro machines~\cite{Barry2006}. 

To better understand bacterial taxis that is driven by hydrodynamically bundled flagellar filaments as well as to assemble flagella based soft-materials it is essential to elucidate the structure as well as the intermolecular forces between flagellar filaments~\cite{Trachtenberg1992,Hoshikawa1985}. Using small-angle X-ray scattering (SAXS) we investigate the behavior of L-type straight flagellar filaments isolated from SJW1660 strain, which differs from the wild-type SJW1103 flagellin strain by the point mutation G426A. The high-resolution structure of the flagellar filament was determined in solution. Under osmotic stress the filaments formed bundles. To quantitatively model scattering patterns from flagellar bundles we performed Monte Carlo simulations that accounted for the effect of thermal fluctuations on the arrangement of the filaments within the bundles. The line-shape of the structure-factor correlation peak and the measured osmotic pressure-distance curves were consistent with theoretical predications~\cite{Selinger1991,Strey1997}. These experiments and models allowed us to determine the contributions of the hydration, electrostatic, and elastic interactions to the equation-of-state describing the lateral forces acting between the flagellar filaments within the bundles and the bending stiffness of the filaments. 
 
\section*{MATERIALS AND METHODS}
\subsection*{Experimental}
L-type straight filament SJW1660, with flagellin point mutation G426A, was isolated from a mutant strain of the wild-type SJW1103 purified from \textit{Salmonella enterica} serovar Typhimurium~\cite{maki2010conformational} following previously published protocol~\cite{Barry2006}. Briefly, bacteria was grown to a log-phase, sedimented at $\unit[8000]{g}$ and redispersed in a minimum amount of volume by repeated pipetting with 1 mL pipette. A very dense foamy bacterial solution was vortex mixed at the highest power setting for $\unit[5]{min}$ to separate flagella from the bacterial bodies (Genie 2 Vortex). Subsequently, this suspension was diluted with a buffer and centrifuged at $\unit[8000]{g}$ for $\unit[20]{min}$ to sediment bacterial bodies. The supernatant contained flagellar filaments, which were then concentrated by two centrifugation/resuspension steps at $\unit[100,000]{g}$ for $\unit[1]{hour}$. For all experiments the flagellar were resuspended in  $\unit[100]{mM}$ \ce{NaCl} and $\unit[10]{mM}$ \ce{K2HPO4}.  We used polyethylene glycol (PEG) of molecular weight $\unit[20]{kDa}$ (purchased from Sigma-Aldrich and used as received) to apply osmotic stress to the flagellar filaments and induce their bundle formation. Osmotic stress samples were prepared by mixing PEG and flagella filament solutions, as described elsewhere~\cite{Steiner2012,needleman2004synchrotron,needleman2005radial, Szekely2012}. The osmotic pressure, $\Pi$, of each polymer solution was measured using a vapor pressure osmometer (Vapro 5520, Wescor, Inc.) and verified against the well-established \cite{Cohen2009} expression: $\log\pi  = a + b\times (wt\%)^c$, where $a = 1.57$, $b = 2.75$, and $c = 0.21$ . The structural changes at each pressure were measured by SAXS.

\subsection*{Solution X-ray Scattering Data Analysis}
Solution small angle x-ray scattering (SAXS) measurements were performed using an in-house setup, or at the ID02 beamline at the ESRF synchrotron \cite{Nadler2011}. To analyze the data, we simulated the real-space structure of the flagella bundle and the interactions between neighboring filaments, and calculated its scattering intensity, $I$, as a function of $q$, which is the magnitude of the momentum transfer vector $\vec{q}$ (or scattering vector) \cite{Ben-Nun2010, Székely2010, BenNun2016132,IJCH:IJCH201500037,doi:10.1021/acs.jcim.6b00159}. By comparing the simulation results with the data we determined the structural parameters and physical properties of the bundles. 

\begin{figure}[h]
\centerline{\includegraphics[width=3.25in]{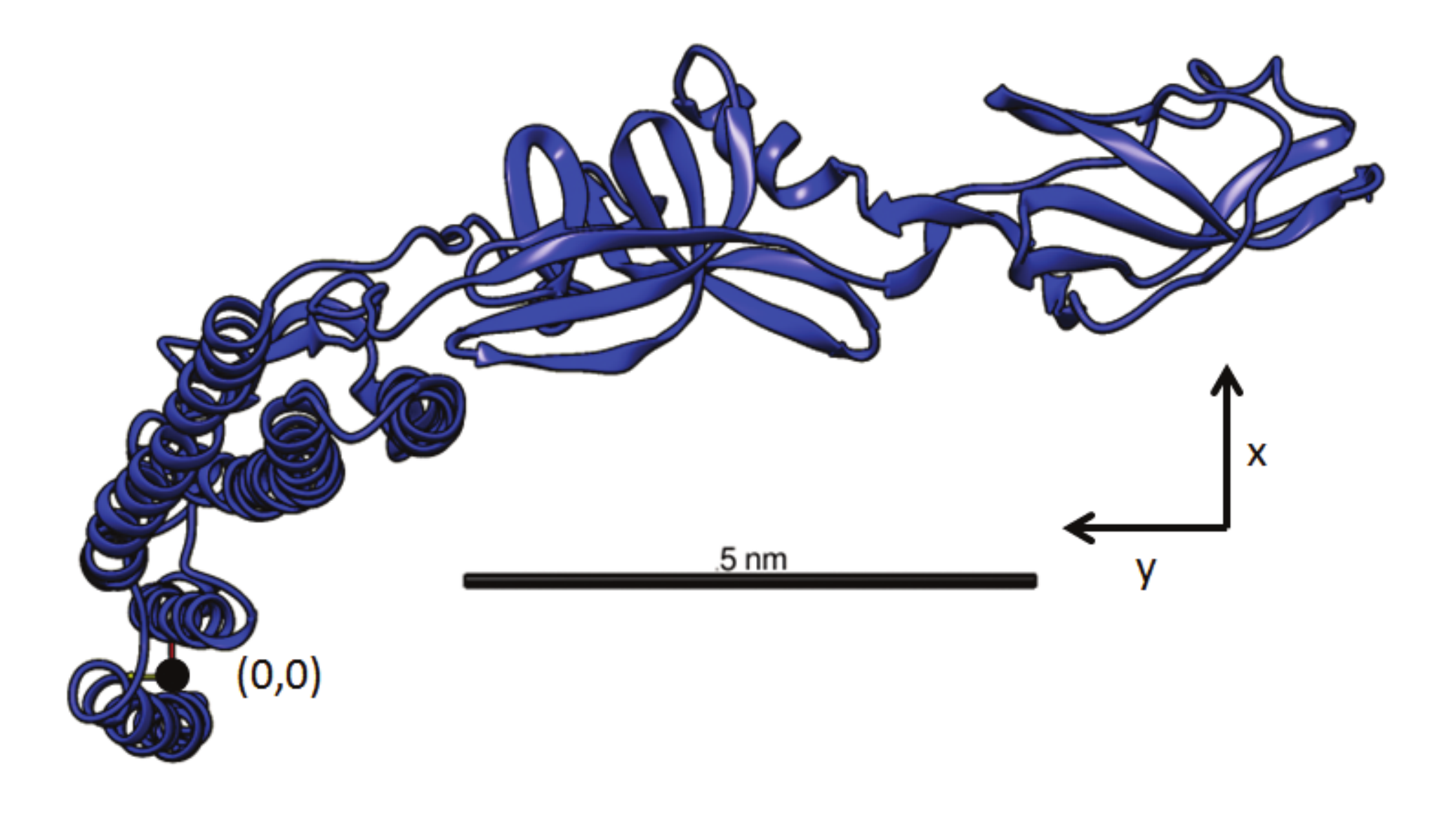}}
\caption{A 3D representation of the flagellin monomer, based on PDB 3A5X \cite{maki2010conformational}. The coordinate system used in our computation model is indicated by the origin, the $x$ and $y$ axes, and the scale bar.}
\label{ltypeaxis}
\end{figure}

The initial structural-parameters guess of the model was taken from electron cryo-microscopy data~\cite{Mimori-Kiyosue1996}. The atomic structure (at $4\AA$ resolution) of a L-type flagellin monomer was taken from the protein data bank (PDB)/3A5X \cite{maki2010conformational,Yonekura2003} and placed in a Cartesian coordinate system (Fig.~\ref{ltypeaxis}). The origin was not placed at the subunit center of mass, but rather in between the two $\alpha$-helices at the $z$ axis which points along the filament's long axis. This choice allowed a simpler relation between translations and rotations of the monomer with the filament axes. In particular, the scattering amplitude of atom $i$ was calculated using the IUCR atomic form-factor:

\begin{equation}
\ f_{i}^{0}\left(q\right)=\sum_{j=1}^{4}a_{j}\cdot \exp{\left(-b_{j}\left(\frac{\left|\vec{q}\right|}{4\pi}\right)^{2}\right)}+c,
\end{equation}

\noindent where $a_{j},b_{j}$, and $c$ are the Cromer-Mann coefficients, given in Table 2.2B of the International Tables for X-ray Crystallography \cite{hamilton1974international} and its subsequent corrections \cite{marsh1983corrections}.

The scattering amplitude of the entire flagellin monomer is given by:

\begin{equation}
\
F_{\text{Monomer}}\left(\vec{q}\right)=\sum_j f_{j}^{0}\left(q\right)\cdot \exp{\left({i\vec{q}\cdot\vec{r}_{j}}\right),}
\label{monomerFF}
\end{equation}

\noindent where $\vec{r}_{j}$ is the location of the $j$th atom in the monomer with respect to the origin, and $\vec{q}$ is the  momentum transfer vector in reciprocal-space. To account for the contribution of the solvent, its displaced volume should be estimated~\cite{svergun1995crysol,Koutsioubas:kk5146,SchneidmanDuhovny2013962}. A uniform sphere (dummy atom) with a mean solvent electron density $\rho_0$ and atomic radius $r^s_j$ could have been placed at the center of each atom $j$ in the PDB file. This approach, however, may generate errors at low $q$ \cite{fraser1978improved}. Therefore,  the uniform spheres were replaced by spheres with Gaussian electron density profiles \cite{fraser1978improved}:
\begin{equation*}
\rho_j\left(\vec{r}\right)=\rho_0\exp\left[-\left(\frac{\vec{r}}{r^s_j}\right)^2\right]
\end{equation*}
\noindent where $\rho_0$ is the mean electron density of the solvent ($\rho_0^{\text{water}}=\unit[333]{\nicefrac{e}{nm^3}}$), and the $r^s_j$ radii were published previously~\cite{svergun1995crysol}. When absent, empirical radii \cite{slater1964atomic} were used. The scattering amplitude contribution of the Gaussian dummy atom is:
\begin{gather*}
F_j\left(\vec q\right) =\\ \int_{0}^{2\pi}d\phi_r\int_{0}^{\pi}d\theta_r\int_{0}^{\infty}\rho_0\exp\left[-\left(\nicefrac{r}{r^s_j}\right)^2\right]\exp\left[i \vec{q} \cdot\vec{r}\right]r^2 \sin\theta_r
dr  
\end{gather*}

\begin{figure}[h]
\centerline{\includegraphics[width=3.25in]{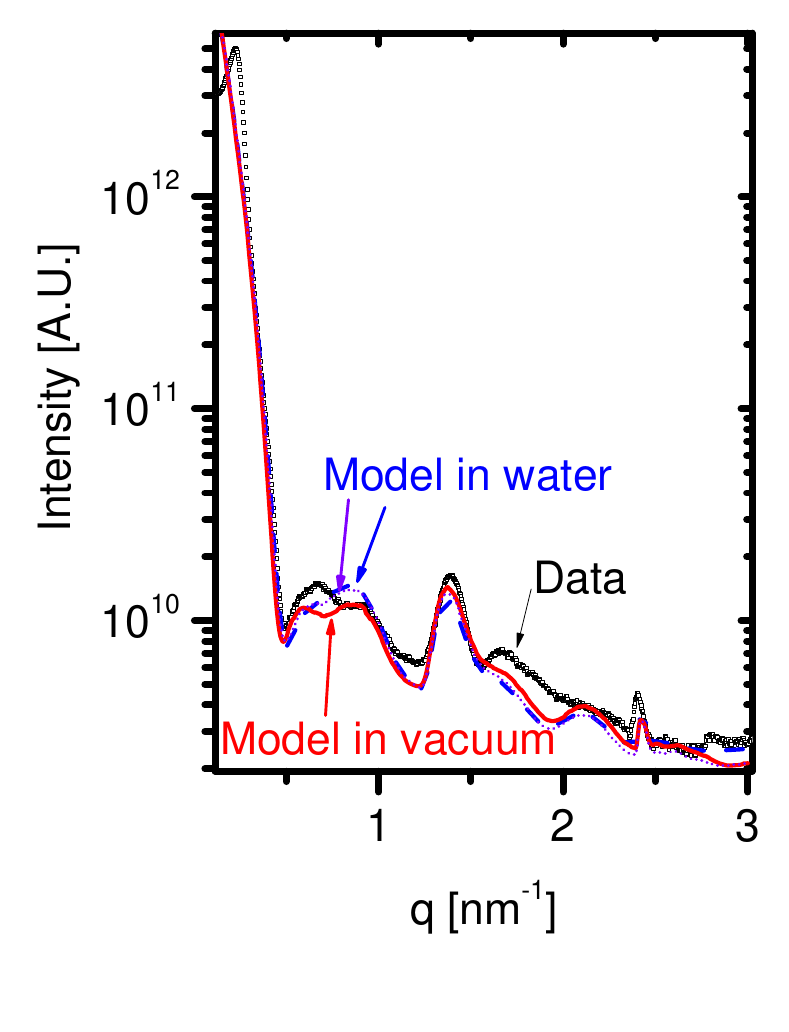}}
\caption{Radially integrated scattering intensity from SJW1660 strain (empty square symbols) and the computed form-factor of a flagella filament, averaged over all
  orientations in $\vec{q}$-space, using Eqs.\ref{monomerFF} and \ref{Intensity} (red solid curve) and after taking into account the contribution of the displaced solvent, using Eq. \ref{Intensity} and Eq. \ref{bsmonomer} with solvent mean electron density of $\rho_0 =\unit[333]{\nicefrac{e}{nm^3}}$ (blue broken curve) or with $\rho_0 = \unit[303]{\nicefrac{e}{nm^3}}$ (dotten violet curve). }
\label{SoventSubtractedFF}
\end{figure}

\noindent The result depends on the radius and $q$, owing to the spherical symmetry, and is given by:
\begin{equation}
F_j\left(q\right) = 
		\rho_{0}\pi^{\frac{3}{2}}\left(r^s_j\right)^3 \exp\left[-\left(\frac{r^s_j \cdot q}{2}\right)^2 \right].
\label{dummysolvent}
\end{equation}
In this approach, the overall excluded volume $V_{\text{ex}}^{\text{Gaussian}}$ is $\pi^{\frac{3}{2}}\left(r^s_j\right)^3$, and is larger by a factor of $ \frac{3\sqrt{\pi}}{4} \approx 1.33$ than the volume of the uniform sphere, $V_{\text{ex}}^{\text{Uniform}}=\frac{4\pi}{3}\left(r^s_j\right)^3$, in agreement with previous work~\cite{fraser1978improved}.
To better fit the data the value of the mean electron density, $\rho_0$ ,was adjusted to some extent (Figure S\ref{SoventSubtractedFF}).    
When the solvent contribution was taken into account, the scattering amplitude from a monomer became: 
\begin{equation}
\
F_{\text{Monomer}}\left(\vec{q}\right)=\sum_j \left[f_{j}^{0}\left(q\right)-F_j(q)\right]\cdot \exp{\left({i\vec{q}\cdot\vec{r}_{j}}\right)}.
\label{bsmonomer}
\end{equation}

To describe the entire filament, we first translated the $i$-th monomer, with respect to its origin reference-point (Fig. \ref{ltypeaxis}), by the translation vector $\vec{R_i}\left(x_i, y_i, z_i\right)$. The monomer was then rotated by its Tait-Bryan \cite{roberson1988dynamics} rotation angles, $\alpha_i, \beta_i, \gamma_i$, around the $x,y,z$ axes, respectively, using the rotation matrix: 

\begin{gather*}
\mathbf{
A\left(\alpha,\beta,\gamma\right)=}\\
\begin{bmatrix}\cos \beta\cos \gamma & -\cos \beta\sin \gamma & \sin \beta\\
\cos \alpha\sin \gamma+\cos \gamma\sin \alpha\sin \beta & \cos \alpha\cos \gamma-\sin \alpha\sin \beta\sin \gamma & -\cos \beta\sin \alpha\\
\sin \alpha\sin \gamma-\cos \alpha\cos \gamma\sin \beta & \cos \gamma\sin \alpha+\cos \alpha\sin \beta\sin \gamma & \cos \alpha\cos \beta
\end{bmatrix}.
\end{gather*}

\begin{figure}[h]
\centerline{\includegraphics[width=3.25in]{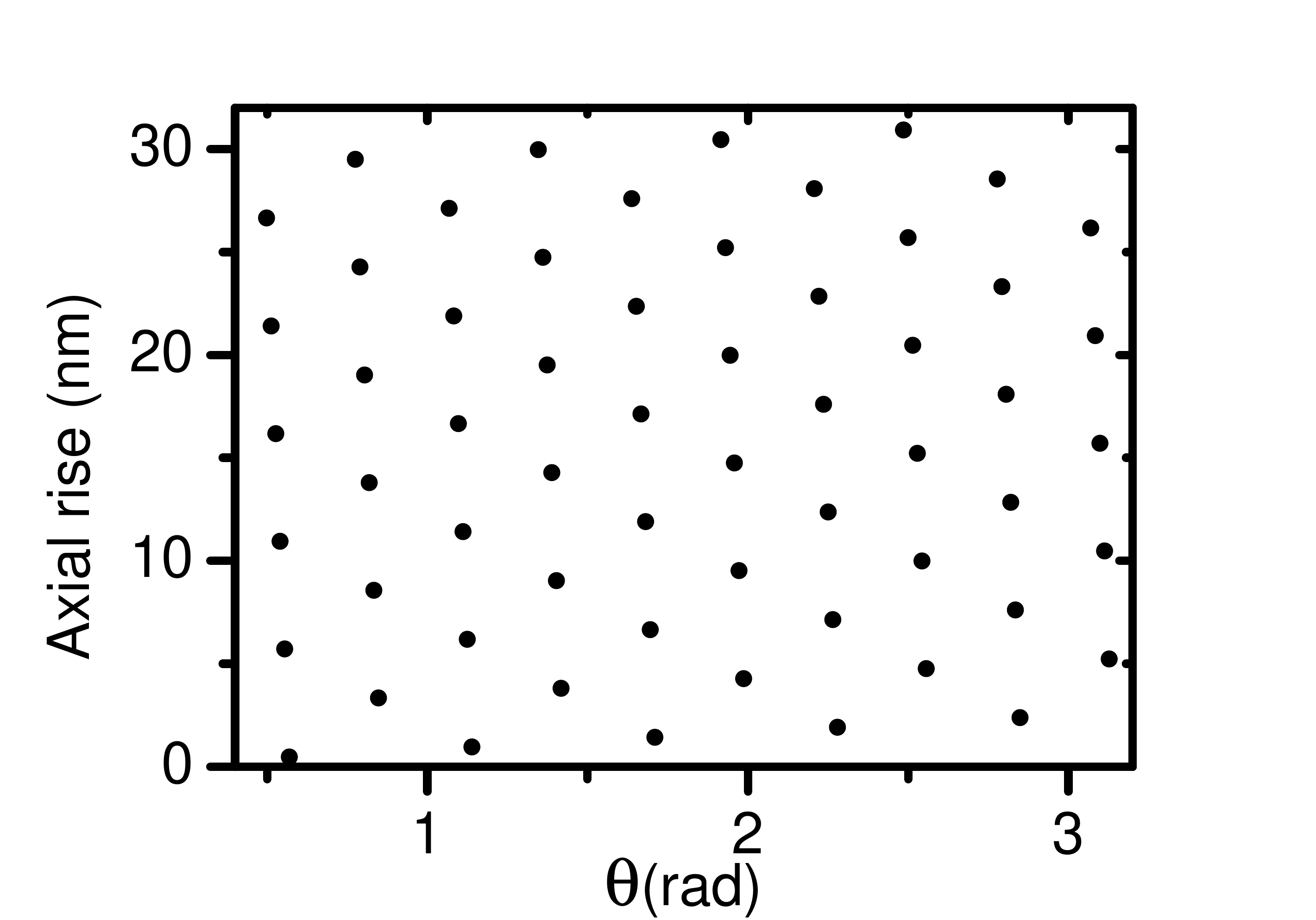}}
\caption{A projection of the helical lattice showing the monomer axial rise ($0.4745\:\textrm{nm}$) and the packing arrangement at the radius of the reference point ($R = 2.42\:\textrm{nm}$). }
\label{Hellical_lattice}
\end{figure}

\noindent The location and orientation of the $i$-th subunit were described as follows:
\begin{equation}
\begin{split}
&\alpha_i=\beta_i=0,\:\;
\gamma_i=i\cdot\frac{2\pi}{NU}\;\:
\\&x_i=R\cdot cos(\gamma_i),\:\;
y_i=R\cdot sin(\gamma_i),\:\;
z_i=i\cdot \frac{P}{NU}
\end{split}
\label{SUplace}
\end{equation}
\noindent where $R$ is the radius of the reference point, $P$ is the two-pitch distance and $NU$ is the number of subunits in a two-pitch turn.
Our model assumes that thermal fluctuations within each flagellum are negligible. Figure \ref{Hellical_lattice} shows a $2D$ projection of the helical lattice of the filament.  

Using the reciprocal grid (RG) algorithm \cite{doi:10.1021/acs.jcim.6b00159}, the flagella scattering form-factor, $FF$, was numerically calculated:  
\begin{equation}
FF(\vec{q})=\sum_{i=1}^{n}F_{\text{Monomer}}\left(\mathbf{A}_{i}^{-1}\left(\alpha_i,\beta_i,\gamma_i\right)\cdot\vec{q}\right)\cdot\exp\left(i\vec{q}\cdot\vec{R_{i}}\right)
\label{FF}
\end{equation}
and orientationally averaged to give the solution scattering intensity:
\begin{equation}
I\left(q\right)=\frac{\int_{0}^{2\pi}d\phi_{q}\int_{0}^{\pi}\left|FF(q,\theta_q,\phi_q)\right|^{2}\sin\theta_{q}d\theta_{q}}{\int_{0}^{2\pi}d\phi_{q}\int_{0}^{\pi}\sin\theta_{q}d\theta_{q}}.
\label{Intensity}
\end{equation}
where  $\mathbf{A}_{i}$ and $\vec{R}_{i}$ are the rotation matrix and the translation vector of the $i$-th monomer, respectively. 

\begin{figure}[h]
\centerline{\includegraphics[width=3.25in]{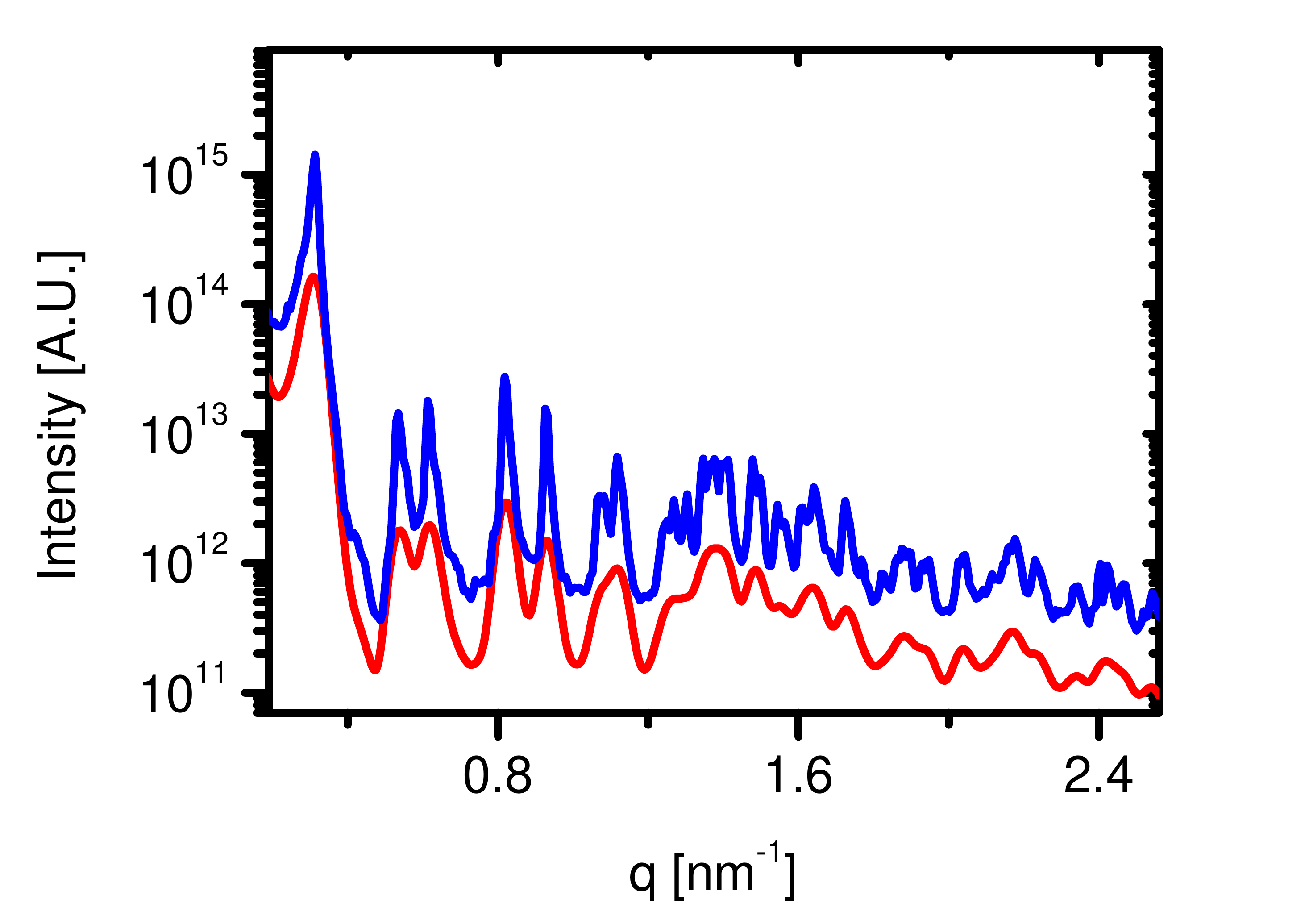}}
\caption{The calculated intensity (as a function of $q$) of a $15\times15$ bundle of filaments in a hexagonal lattice with no fluctuations(blue curve). The red curve results from the a convolution between the blue curve and a Gaussian resolution function with a standard deviation of $\sigma=\unit[0.03]{nm^{-1}}$.}
\label{SF0itr}
\end{figure}

Osmotic stress exerted by non-adsorbing polymers induced bundling of straight flagellar filaments. The scattering intensity owing to the packing of the flagellar filaments was computed by using the form-factor of a single filament, $FF$, as a unit cell. The form-factor was multiplied by a structure-factor (lattice sum), $SF\left(\vec{q}\right)$, of a $2D$ lattice, and orientationally averaged  ($FF\left(\vec{q}\right)\cdot SF\left(\vec{q}\right)$ ) in $\vec{q}$-space. We assumed a perfect hexagonal $2D$ lattice as our starting point (Figure S\ref{SF0itr}). This assumption is equivalent to assuming that the chains are very stiff. The fitting parameters of a hexagonal lattice were the spacing between the centers of neighboring filaments, $a$, and the domain size (the distance over which two lattice points maintain positional correlation). These parameters determined the position and width of the correlation peaks in the scattering intensity. Since samples were in solution at room temperature, the lattice exhibited significant thermal fluctuations, which washed away the sharp peaks of the structure factor (Figure S\ref{SF0itr}). 

In real space, a finite $2D$ lattice, at zero temperature, is described by:
\begin{equation}
SF_r(\vec{r})=\delta(z)\sum_i^M\delta(x-x_i)\delta(y-y_i)
\end{equation}
\noindent where $\vec{r}_i=(x_i,y_i,z_i)$ is the location of the $i$-th point in a lattice with $M$ unit cells. At a finite temperature, thermal fluctuations work against the intermolecular forces, affecting the lattice structure. Assuming a harmonic potential between nearest neighbors, we calculated the pairwise energetic cost, $\Delta E_i$, for a small
displacement, $\Delta\vec{r}$, of the $i$-th lattice point from its mean location, $\vec{r}_i$, at a given temperature:
\begin{equation}
3\Delta
E_i=\frac{1}{2}\kappa\cdot\sum_{j\in nn}\left[(\vec{r}_{\perp i}+\Delta\vec{r}_{\perp}-\vec{r}_{\perp j})^2-(\vec{r}_{\perp i}-\vec{r}_{\perp j})^2\right]
\label{deltaE}
\end{equation}

\noindent Here $\vec{r}_{\perp i}=\left(x_i,y_i,0\right)$ and $nn$ denotes nearest neighbors and $\kappa$ is the lattice elastic constant between neighbors. The factor of three accounts for the fact that, on average, each chain has six neighbors and the interaction is shared between the interacting pairs \cite{ben2013entropy}. The probability of a deviation $\Delta E_i$ in the energy is:
\begin{equation}
P_i\left(\Delta
 E_i\right)\sim \exp\left(-\frac{\Delta E_i}{k_BT}\right)
 \label{Ptest}
\end{equation}
 
\noindent where $k_B$ is Boltzmann constant and $T$ is the absolute temperature.  
To estimate the effect of thermal fluctuations, we performed Monte-Carlo simulations. In each iteration, we tested the probability of a random displacement at a random lattice-point against a random number between $0$ and $1$. If the random number was smaller than the calculated probability, $P_i\left(\Delta E_i\right)$, the displacement was accepted. Repeating this process for ca. $10^9$ iterations, using periodic boundary conditions, converged into a stable, slightly (depending on the value of $\kappa$) disordered $2D$ hexagonal lattice. Figure S\ref{SFcomp} shows how the value of $\kappa$ affected the calculated intensity in this model. 

In real space, the total electron density is a convolution of the electron density of a filament, $\rho_{\textrm{filament}}(\vec{r})$, and the $2D$ bundle lattice, $SF_r\left(\vec{r}\right)$. In reciprocal space, the convolution becomes a multiplication, hence, the total scattering amplitude is:
\begin{equation}
F\left(\vec{q}\right)=FF\left(\vec{q}\right)\cdot SF\left(\vec{q}\right),
\label{SA}
\end{equation}  
where 
\begin{equation}
SF\left(\vec{q}\right)=\sum_{i=1}^{M}\exp\left(i\vec{q}\cdot\vec{r}_{i}\right).
\label{Sq}
\end{equation}
To obtain the scattering intensity,  $I_{\text{bundle}}\left(q\right)$, we calculated the square of the scattering amplitude, $\left|F(q,\theta_q,\phi_q)\right|^{2}$, and averaged over all the orientations in reciprocal space ($\theta_q$ and $\phi_q$), as in Eq. \ref{Intensity} \cite{doi:10.1021/acs.jcim.6b00159}. $I_{\text{bundle}}\left(q\right)$ was then compared with the experimental SAXS data.

\section* {Results and Discussion}
\subsection*{Scattering from flagellar suspensions} 
\begin{figure}[h]
\centerline{\includegraphics[width=3.25in]{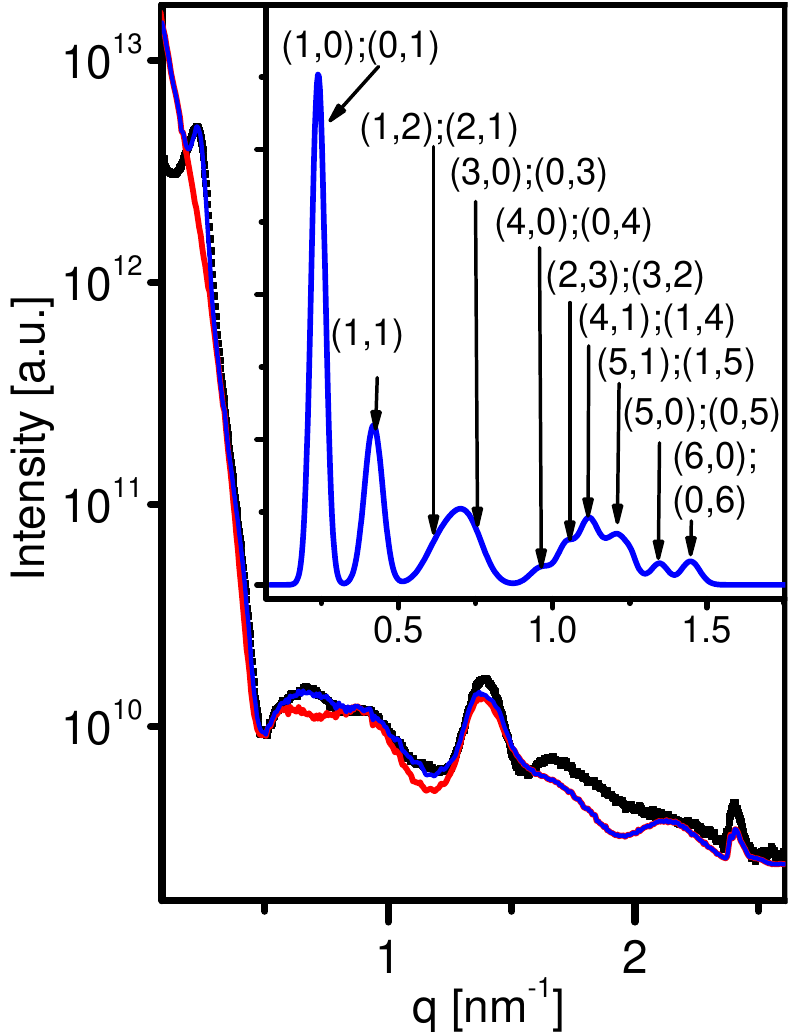}}
\caption{Radially integrated scattering intensity as a function of $q$ (the magnitude of the momentum transfer vector, $\vec{q}$), from isotropic solution of SJW1660 strain (black solid square symbols). The form-factor (red curve) of a flagella filament was computed using Eqs. \ref{monomerFF} and \ref{Intensity} and fitted to experimental data yielding the following essential flagellar structural parameters: filament diameter $D=\unit[23.1]{nm}$, two turn pitch, $p=\unit[5.2]{nm}$, $10.96$ flagellin subunits per two turn pitch, and a radius of the reference point, $R = \unit[2.42]{nm}$. The blue curve corresponds to same form-factor when multiplied by the structure-factor of a $2D$ hexagonal phase with a lattice constant, $a$ of $\unit[30.0]{nm}$. The structure-factor and its peak indexes are shown at the inset.}
\label{FF1660}
\end{figure}

The X-ray scattering $2D$ pattern from a dense solution of straight flagella, isolated from strain SJW 1660, was azimuthally averaged yielding experimental scattering intensity curve (Fig. \ref{FF1660}). The experimental data were compared with a scattering curve computed from a flagellar model in which the atomic structure of flagellin monomers (Fig.~\ref{ltypeaxis}) were arranged on a  flagellar one-start left-handed helical structure. The relevant microscopic parameters in Eq. \ref{SUplace} ($R$, $P$, and $NU$) were varied to obtain the best fit of our theoretical computed curve to the experimental data. During this procedure the atomic structure of the flagellin subunit, obtained from the PDB file $3A5X$ \cite{maki2010conformational}, was preserved and the subunits were not allowed to overlap. 

\begin{figure}[h]
	\begin{center}
    \includegraphics[width=3.25in] {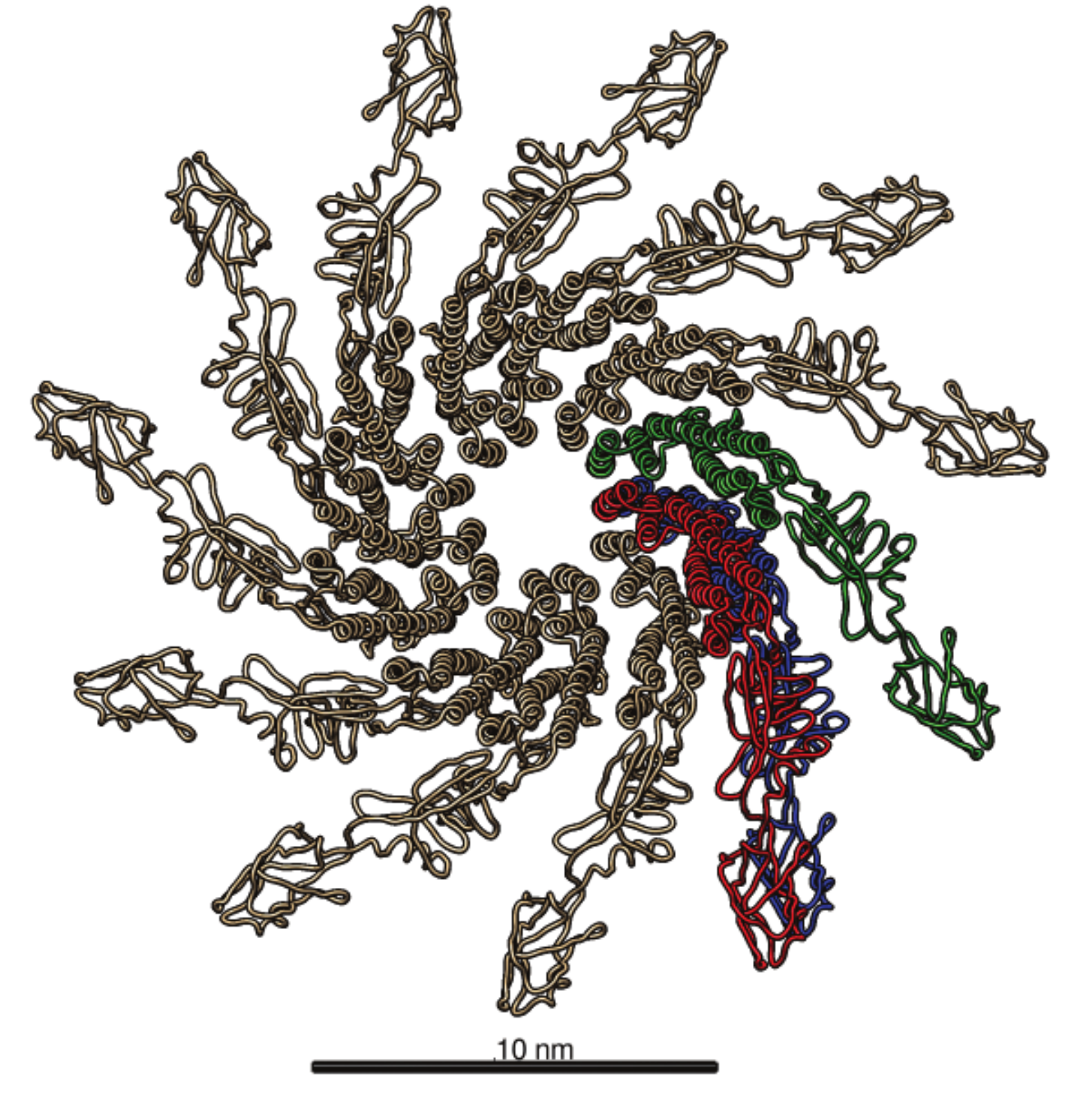}
    \hspace{3pt}
    \includegraphics[scale=0.3]{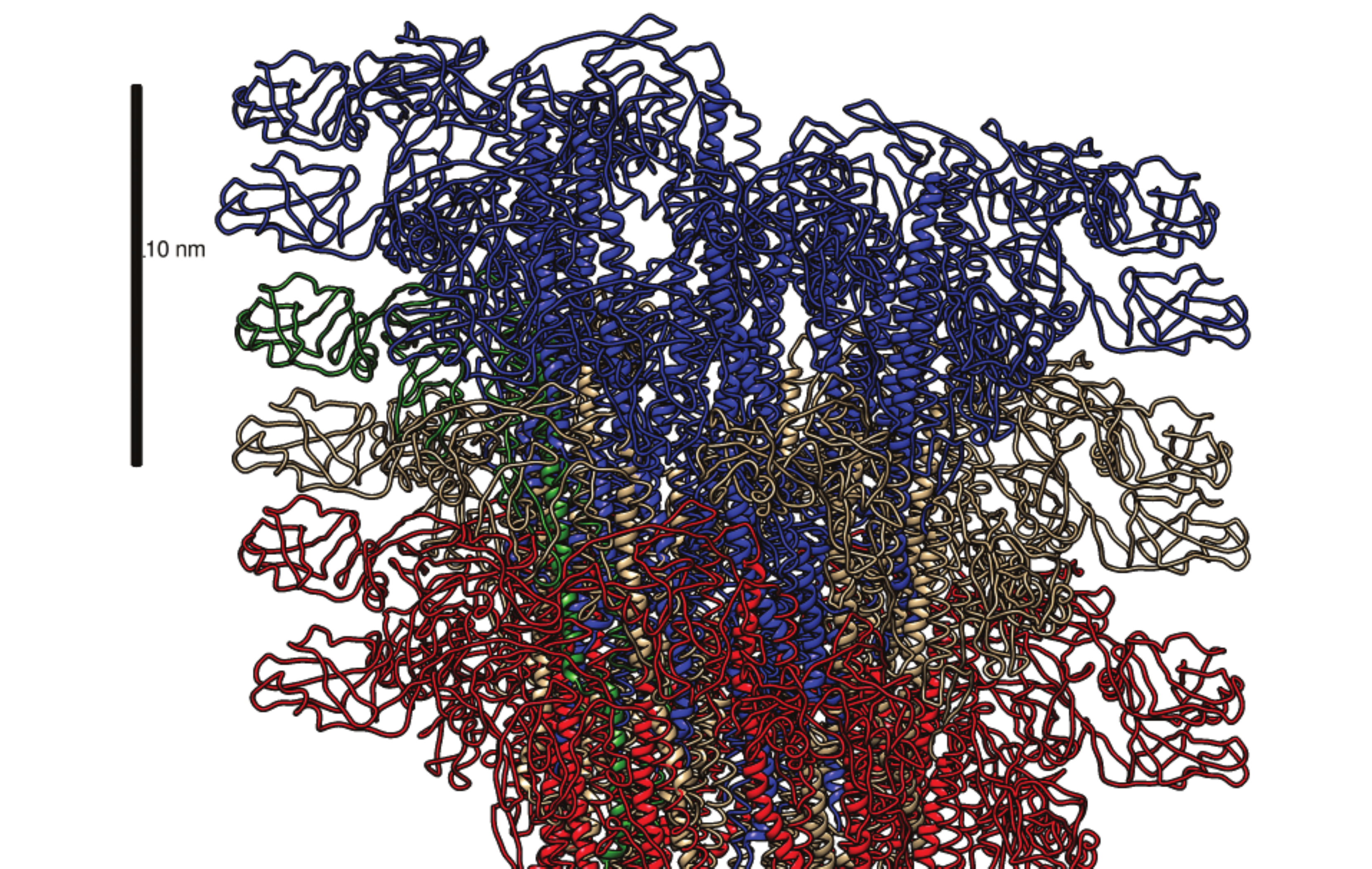}
    \caption{{\bf a)} $3D$ model of a two turn pitch (11 flagellin subunits) viewed along the filaments long-axis. The blue subunit is the first monomer (at $z=0$) and the red subunit is the monomer unit of the second two turn pitch, a small deviation to the left of the positioning can be seen, see also Figure  \ref{Hellical_lattice}. {\bf b)} A structure of a $3$ pitch filament viewed from the side. }
	\label{3Dmodel}
	\end{center}
\end{figure}

In our experiments, the average length of flagellar filaments was ~4 $\unit{\mu m}$~\cite{Barry2006}. The SAXS measurements, however, were insensitive to objects longer than a few hundred $\unit{nm}$, hence in our theoretical computations, the filament length, $L$ was fixed at $\unit[300]{nm}$ (in other words, the filament contained $n = 660$ monomers). Models with longer filaments required more computational resources and did not change the scattering intensity profile or better fit the data, in the $q$-range of our data. Quantitative comparison of experimental measurements to the theoretical model revealed a filament diameter, $D$, of $\unit[23.1]{nm}$, a two turn pitch, $P$, of, $\unit[5.2]{nm}$, $NU=10.96$ flagellin subunits per two turn pitch, and a radius of the reference point, $R$, of $\unit[2.42]{nm}$ (Fig. \ref{3Dmodel}). These values are consistent with electron cryo-microscopy and X-ray fiber diffraction data ($D=\unit[23-24]{nm}$~\cite{Yonekura2003,hasegawa1998structure}, $P=\unit[5.27]{nm}$, $NU=11.26$~\cite{maki2010conformational, hasegawa1998structure}, and $R=\unit[2.5]{nm}$ ~\cite{maki2010conformational}).

Using Gaussian dummy atoms (Eqs. \ref {bsmonomer} and \ref{Intensity}) to account for the contribution of the solvent with reasonable values for the solvent electron density, $\rho_0$ (in Eq. \ref{dummysolvent}), did not significantly improve the fit of our form-factor model to the experimental data (Figure S\ref{SoventSubtractedFF}). The contribution of the solvent was therefore not computed in subsequent models (in other words, Eqs. \ref{monomerFF} and \ref{Intensity} were used). By varying the model parameters, we determined the effect of each parameter on the locations and magnitudes of various features in the calculated intensity. The first minimum was mainly controlled by the flagellar filament diameter  (see Figure S\ref{FFradius}), the peak at $q=\unit[1.4]{nm^{-1}}$ was closely associated with the helical pitch (see Figure S\ref{FFpitch}), and the peak at $q=\unit[2.4]{nm^{-1}}$ was attributed to the axial rise (Figure S\ref{FFtilt}). The high sensitivity of our model to these structural parameters and their weak interdependency is demonstrated in Figures S\ref{FFradius}, S\ref{FFpitch}, and S\ref{FFtilt}. 

To obtain high signal to noise scattering patterns, we used $\unit[27\pm 0.1]{mg/ml}$ (or $\unit[0.524\pm 0.002]{mM}$) flagellar suspension. Based on the structure of the flagellar filaments, the mean filament volume fraction, $\phi$, in this sample, was $\simeq 0.062$. Previous work has demonstrated that rigid rods form a nematic phase when $\nicefrac{\phi L}{D}>4$. This relationship become quantitatively valid in the Onsager limit in which $L/D >100$ \cite{onsager1949effects, fraden1989isotropic}. Consequently, because our flagellar filaments were longer than $\unit[2]{\mu m}$, and thus satisfied the Onsager criterion, the filament suspension formed nematic liquid crystals. We note that inherently polydispersity of flagellar filaments significantly widens the isotropic-nematic co-existence. This makes it possible that shorter filaments partitioned into an isotropic phase ~\cite{wensink2003isotropic}.  Furthermore, rigorous analysis would have to account for contribution of electrostatic repulsion which leads to effective diameter that can be significantly larger than the bare one. The high ionic strength of our suspension, however, significantly reduced this contribution ~\cite{stroobants1986effect}.

While the majority of the measured scattering pattern from flagellar filaments is owing to the form factor, the signal also contained weak structure-factor correlation peaks, at $q_{(1,0)}=\unit[0.24]{nm^{-1}}$ and its higher harmonics (Figure \ref{FF1660}). The presence of these peaks suggest that a fraction of the filaments within our sample formed hexagonal bundles with lattice constant, $a$, of $\unit[30.0]{nm}$ (inset to Fig.~\ref{FF1660}). Within such bundles the volume per chain is $\nicefrac{\sqrt{3}La^2}{2}$, where $L$ is the mean filament length. The volume of a chain is $\nicefrac{L\pi D^2}{4}$  hence the volume fraction of the filaments in our lattice, given by the ratio of the two, $\phi=\nicefrac{\pi D^2}{2\sqrt{3}a^2} $, was $\simeq 0.54$. The average volume fraction of the filaments, however, was $\simeq 0.062$, suggesting that a low density nematic liquid crystal coexisted with a low fraction of filaments that formed high density hexagonal bundles.

Lindemann stability criterion asserts that the root-mean-squared displacement,  $\langle\left|\vec{u}\right|^2\rangle^{\nicefrac{1}{2}}$, in a lattice should be small compared with the lattice constant ($<0.1a$). In a lattice with purely steric interactions \cite{Selinger1991}:
\begin{equation}
\langle\left|\vec{u}\right|^2\rangle^{\nicefrac{1}{2}}=3^{\nicefrac{1}{2}}\left(a-D\right).
\label{msd}
\end{equation}
The lattice is expected to melt when $a_{\text{max}} \approx 1.2D$, which in the case of our flagella filaments corresponds to $a\approx \unit[28]{nm}$ \cite{Selinger1991}. For the filaments in the hexagonal phase, $a$ was $\approx1.3D$, suggesting that the van der Waal attractive interactions between filaments are not negligible and stabilize the bundle structure. The structure-factor peaks were relatively wide (FWHM of $\approx\unit[0.06]{nm^{-1}}$), suggesting that hexagonal bundles have small lateral dimensions. Applying Warren's approximation revealed that, on average, there were only $\sim 3$ filaments that maintained positional correlation in the lattice~\cite{warren1941x}. 

\subsection*{Measuring equation of state for flagellar solutions using osmotic stress technique} 
To induce large bundle formation, we applied osmotic stress to the flagellar filaments by adding increasing concentrations of an inert polymer (PEG, $M_w=\unit[20]{kDa}$) to the flagellin solution. We then determined the structure of flagellar filament bundles and the interactions between filaments in the bundles. To obtain the mean interfilament lateral separation, $a$, and the average coherence-length, along which the positional order of filaments within the bundle is maintained, we modeled the scattering from a filamentous bundle, and compared with experimental measurements. We multiplied the single filament form-factor (Eq. \ref{FF}) by a hexagonal lattice-sum (Eqs. \ref{SA} and \ref{Sq}) and orientation-averaged the product in $\vec{q}$-space (Eq. \ref{Intensity}). To take into account the effect of thermal fluctuations, we assumed a harmonic potential between nearest filament neighbors and calculated the energetic cost of random small displacements in the hexagonal lattice (Eq. \ref{deltaE}). We performed Monte Carlo simulation (using Eq. \ref{Ptest}) that equilibrated the lattice structure. A lattice of $40 \times 40$ and $\kappa = \unit[1.2]{mN\cdot m^{-1}} = \unit[0.29]{k_BT\cdot nm^{-2}}$ ($\kappa$ is defined in Eq. \ref{deltaE}), were kept constant. Based on the locations of the filament centers that were obtained from the simulations, we calculated the structure-factor (Eq. \ref{Sq}) and$I_{\text{bundle}}(q)$ (Eq. \ref{Intensity}), and compared these predictions with experimental SAXS data. 

\begin{figure}[h]
\centerline{\includegraphics[width=3.25in]{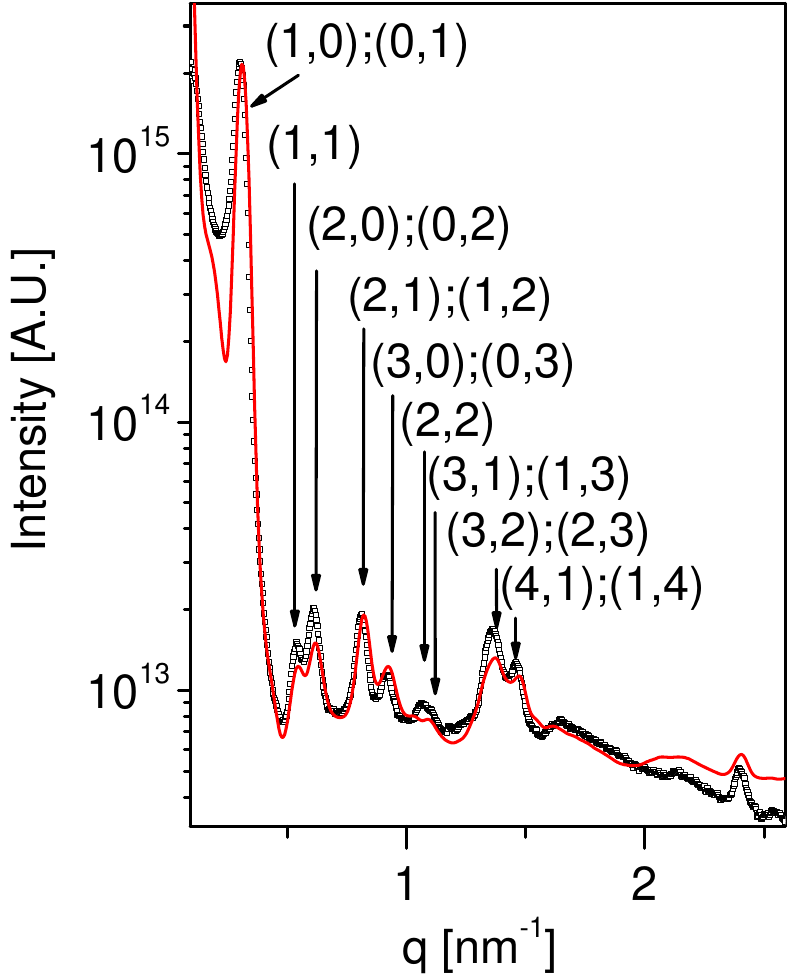}}
\caption{Scattering from a suspension of bundled flagellar filaments. The experimental measurements (black squares) are fitted with our computational hexagonal lattice model (red curve) with a lattice constant  $a=\unit[23.5]{nm}$. Good agreement can be seen over a wide $q$-range both in the location and the magnitude of the peaks, whose indexes are indicated in brackets. The bundles were assembled using $\unit[5]{wt\%}$ PEG $\left(M_w=\unit[20,000]{Da}\right)$ concentration, corresponding to an osmotic pressure of $\unit[47]{kPa}$ \cite{Cohen2009}.}
\label{SF1660}
\end{figure}

\begin{figure}[h]
\centerline{\includegraphics[width=3.25in]{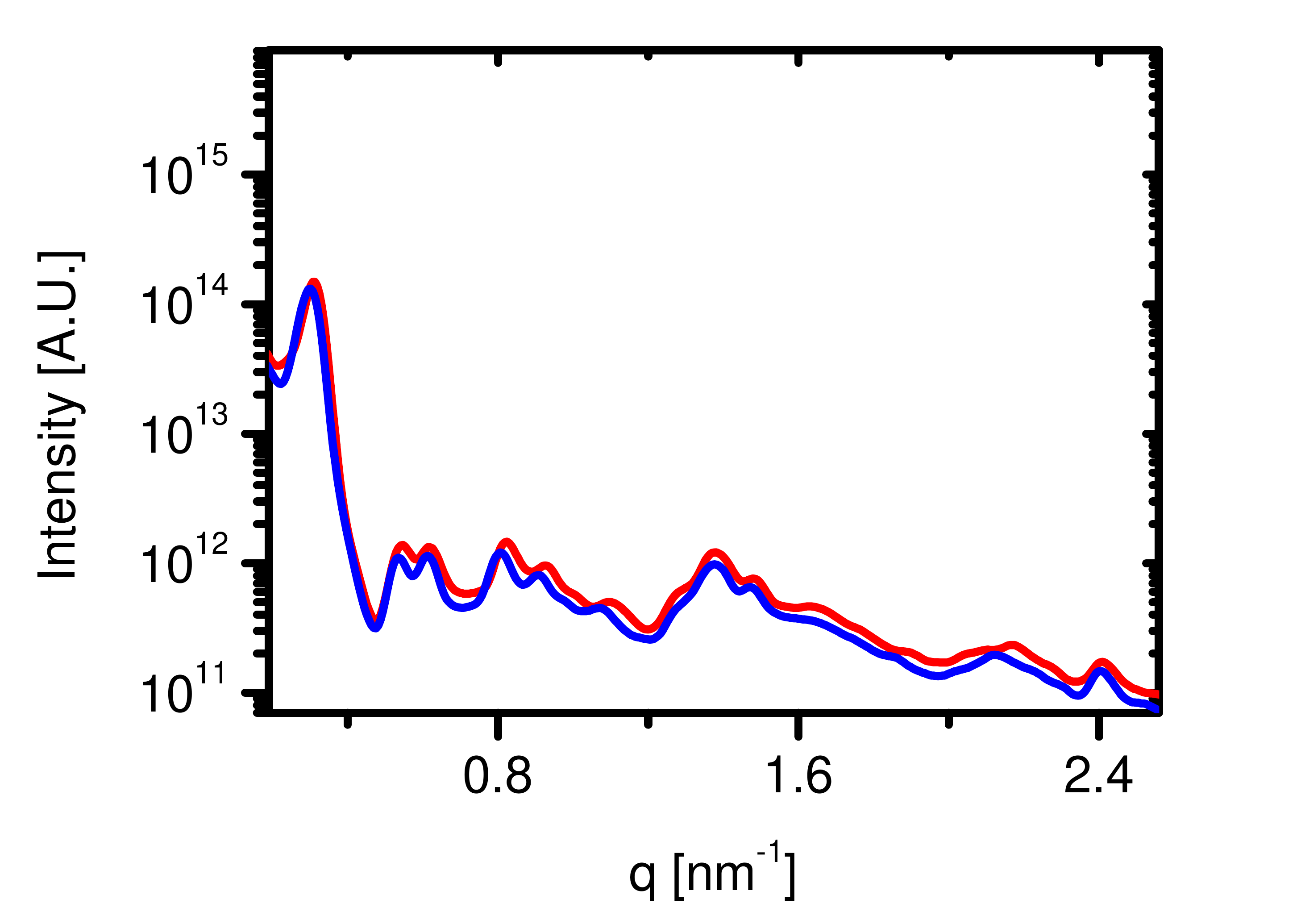}}
\caption{A comparison between the calculated scattering intensities of two models of bundles with a small difference  in their lattice constants. The blue curve is with $a=\unit[23.4]{nm}$ and the red curve with $a=\unit[23.8]{nm}$. As the lattice constant increases the correlation peaks shift to lower $q$ values.}
\label{latticeComp}
\end{figure}

Three lattice parameters affected the scattering intensity,  $I_{\text{bundle}}(q)$. Firstly, the $2D$ hexagonal lattice size, $a$, which determined the locations of the structure-factor correlation-peak centers (Figure S\ref{latticeComp}). Secondly, the lattice coherence-length (i.e. the positional correlation-length of the lattice), which mainly influenced the width of the structure factor correlation-peaks (Figure S\ref{DomainComp}). Finally, the elastic constant, $\kappa$, which affected the intensity and the number of structure-factor peaks (Figure S\ref{SFcomp}). High $\kappa$ values correspond to weaker thermal fluctuations, and hence sharper correlation-peaks.  Our computational model quantitatively fitted the experimental scattering curve over a wide range of $q$ values (Fig.\ref{SF1660}). Results for other osmotic pressures, which show comparable agreements with the computational model are shown in Fig.~S\ref{MorePressures}. From the equipartition theorem and the value of $\kappa$ we can estimate the root-mean-squared displacement of a single flagellar chain confined in the hexagonal lattice to be \cite{ben2013entropy}: $\langle\left|\vec{u}\right|^2\rangle^{\nicefrac{1}{2}}=\sqrt{\nicefrac{k_BT}{\kappa}}\approx \unit[1.85]{nm}$. 

Taking into account hydration repulsion, the bending stiffness, $\kappa_s$, of the flagellar filaments, and the electrostatic interactions between them, the equation-of-states for a bundle of long semi-flexible chains in solution is~\cite{Strey1997}: \begin{equation}
\frac{\partial G}{\partial d}(d)=\frac{\partial H_0}{\partial d}(d)
+ck_BT\kappa_s^{-\frac{1}{4}}\frac{\partial}{\partial
  d}\sqrt[4]{\frac{\partial^2 H_0}{\partial d^2}}
\label{states}
\end{equation}
where G is the free energy, $d = a - D$ is the spacing between filaments,
$\kappa_s$ is the bending stiffness, $H_0$ is:
\begin{equation}
H_0(d)=a_h\frac{e^{-d/\lambda_H}}{\sqrt{d/\lambda_H}}+b\frac{e^{-d/\lambda_D}}{\sqrt{d/\lambda_D}},
\label{H0}
\end{equation}
where $\lambda_H$ and $\lambda_D$ are the hydration and electrostatic screening lengths. The explicit expression of $\frac{\partial G}{\partial d}(d)$ is given in the Supporting Materials ($\sec$ \ref{EofS}). 

In a hexagonal lattice, the relation between the free energy and the osmotic pressure is~\cite{Strey1997}:

\begin{equation}
\Pi(d)=-\frac{1}{\sqrt{3}d}\frac{\partial G}{\partial d}(d).
\label{Gp}
\end{equation}

\begin{figure}[h]
\centerline{\includegraphics[width=3.25in]{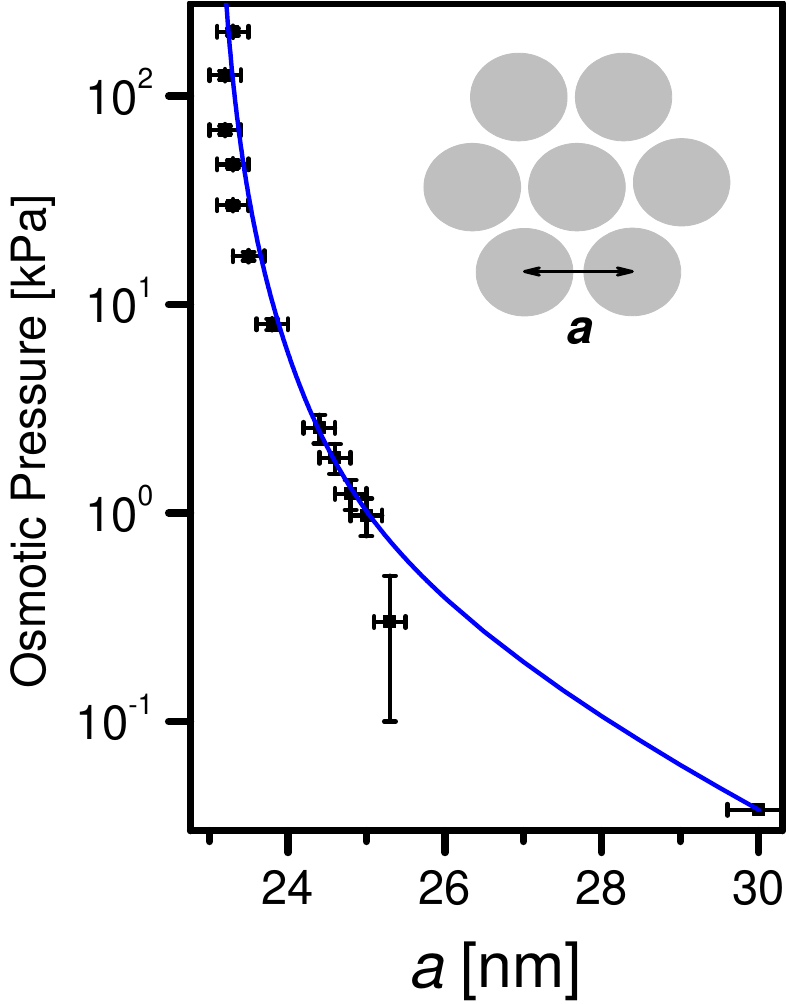}}
\caption{The equation-of-state of bundled straight filaments shows how the lattice constant, $a$, of a hexagonal bundle (see inset for a schematic top view) of straight filaments depends on the applied osmotic pressure. The broken blue curve indicates the fit of the theoretical model (Eq. \ref{states}) to the data.}
\label{OP1660}
\end{figure}

As expected, lateral filament spacing $a$, obtained from fitting the scattering data to our model (as demonstrated in Figure \ref{SF1660}) decreases with increasing osmotic pressure (Figure \ref{OP1660}). The experimental pressure-distance curve could be quantitatively fitted to the theoretical equation-of-state (Eqs. \ref{states},\ref{H0},\ref{Gp}). We fixed the filament hard-core diameter, $D$, to the value obtained from the form-factor analysis $\left(\unit[23]{nm}\right)$ when performing the fit. The fit yielded the following parameters: $a_h=\unit[40\pm 5]{Pa\cdot nm^2}$, $b=\unit[40\pm 5]{Pa\cdot nm^2}$, $c=1.7\pm0.1$, $\kappa_s=\unit[2.9\pm0.4\cdot10^{-15}]{J\cdot nm}$, $\lambda_H=\unit[0.26\pm0.03]{nm}$, and $\lambda_D=\unit[0.82\pm0.03]{nm}$. $\lambda_H$ is close to the expected value of ca. $\unit[3]{\AA}$ \cite{Parsegian1993}. $\lambda_D$ can be calculated  from the calculate ionic strength of the sample: $\unit[0.84]{nm}$ \cite{Jacob2011}. The small difference between the calculated and the measured $\lambda_D$ could be explained by the small amount of \ce{NaOH} that was added to the solution to maintain natural pH. We kept the value of the prefactor $c$ close to unity, as obtained in an earlier study \cite{Strey1997}. $a_h$ and $b$, where first fitted to the high pressure data, where the contribution of the hydration and electrostatic forces should dominate.

The bending stiffness, $\kappa_s$, terms dominate the lower pressure data and is associated with the persistence length of the filaments \cite{Selinger1991}: $P=\nicefrac{k_s}{k_BT}=\unit[700\pm 100]{\mu m}$.  This value is significantly higher than the persistence length of actin ($\unit[18]{\mu m}$), which has a smaller cross section, and comparable to the persistence length of taxol-free microtubule $\left(\unit[700 - 1,500]{\mu m}\right)$, which has a slightly larger cross-section \cite{gittes1993flexural}.  Taxol-stabilized microtubule has, as expected, longer persistence length $\left(\unit[5,200]{\mu m}\right)$ \cite{gittes1993flexural}. The latter values were measured from thermal fluctuations in the shape of the filaments. Note that the persistence length that we found is higher than the value determined from electron micrographs of isolated, negatively stained filaments  $\left(\unit[41]{\mu m}\right)$   \cite{Trachtenberg1992}. Finally, a constant weak negative effective pressure of $\unit[37]{Pa}$ was added to account for the contribution of the van der Waals interaction, which led to the hexagonal phase when no osmotic stress was applied (Fig. \ref{FF1660}). 

\begin{figure}[h]
\centerline{\includegraphics[width=3.25in]{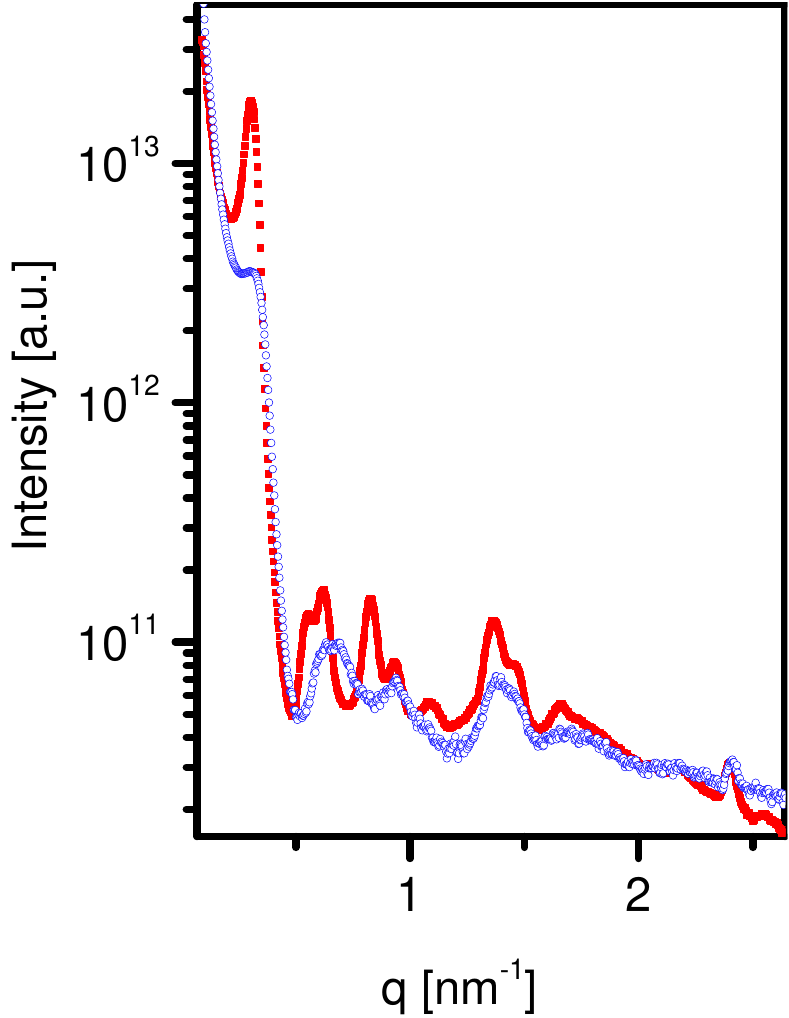}}
\caption{At hight osmotic pressures the scattering profiles exhibit a discontinuous change indicating a structural changes of the constituent flagella.  A comparison between SAXS patterns measured for moderate  $~30\:\textrm{kPa}$ (red solid symbols, hexagonal lattice spacing of $\unit[23.3]{nm}$) and high $\unit[490]{kPa}$ (blue open symbols, hexagonal lattice spacing of $\unit[21]{nm}$) osmotic pressures.}
\label{HP1660}
\end{figure}

Whereas the theoretical model quantitatively fits the data over a wide range of applied osmotic pressures, when the filaments were far apart, the model (Eq. \ref{states}) predicted a slightly more repulsive interaction than measured. At very high pressures ($\unit[490]{kPa}$ or higher), the filaments assumed lattice spacing values, $a$, that were smaller than the unstressed filament diameter. Our data, however, show that at these high pressures the first minimum of the azimuthally integrated scattering curve moved towards higher $q$ values, suggesting that the form-factor has changed owing to deformation of the filaments (Fig. \ref{HP1660}). This change is consistent with a tighter monomer packing resulting in a smaller filament diameter, $D$.
From the structure of the flagellar filaments we can calculate the cross-section geometrical moment of inertia, $I\approx\frac{\pi}{4}\left[\left(\nicefrac{D}{2}\right)^4-R^4\right]= \unit[1.4\times 10^4]{nm^4}$. The filament Young's modulus, $E$, is then given by ~\cite{gittes1993flexural}:
\begin{equation*}
E=\frac{\kappa_s}{I}\approx\unit[0.2]{GPa},
\end{equation*} where we have assumed that $\kappa_s \simeq \unit[700]{\mu m}$.

Semi-flexible flagellar filament chains are confined to an effective ''tube'' within the hexagonal lattice. It has been argued that fluctuations of this type of confined filaments can be described by a single characteristic length scale, which points along the long-axis, $\hat{z}$, direction of the filaments. This length scale is the Odijk deflection length, $\lambda_{\text{def}}$, which is the average displacement between successive collisions along the chain within the confined lattice, and is given by \cite{odijk1983statistics,odijk1986theory}:
 \begin{equation}
 \lambda_{\text{def}}\approx P^{\nicefrac{1}{3}}d^{\nicefrac{2}{3}}.
\label{ld}
 \end{equation}
Equation \ref{ld} is based on scaling theory  and agrees well with MC simulations up to a prefactor of order $2$ ~\cite{DIJKSTRA1993374}. 
The mean fluctuations in the nematic director can then be estimated from the deflection length \cite{Selinger1991}:
\begin{equation}
\langle\left|n_\perp\right|^2 \rangle\approx\left[\frac{d}{\lambda_{\text{def}}}\right]^2.
\end{equation}
\noindent Therefore it follows that the measurement of $d$ directly yields the Odijk deflection length.

\begin{figure}[h]
\centering
	\begin{subfigure}[b]{.55\linewidth}
		
		\includegraphics[width=\linewidth]{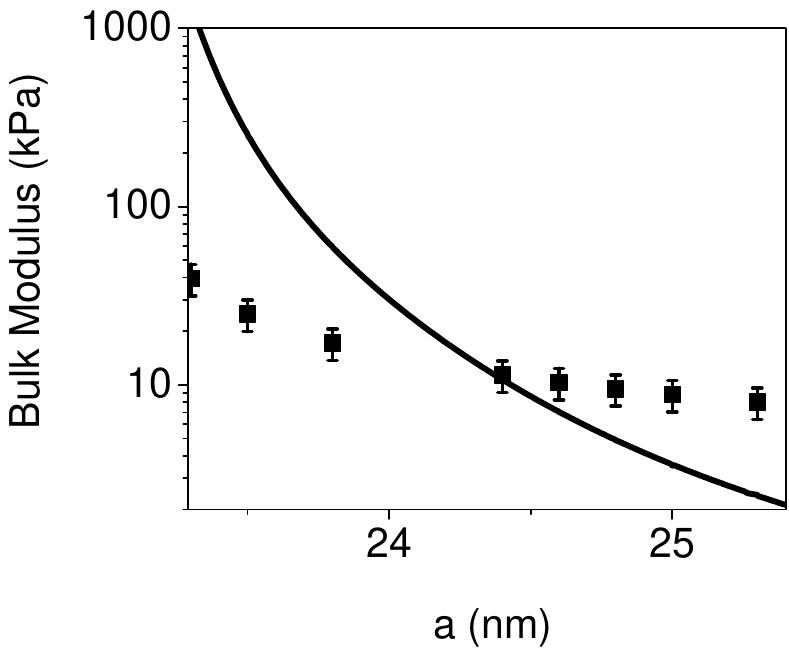}
		\caption{}
		\label{fig:Modulus}
	\end{subfigure} 
	\begin{subfigure}[b]{0.55\linewidth}
		
		\includegraphics[width=\linewidth]{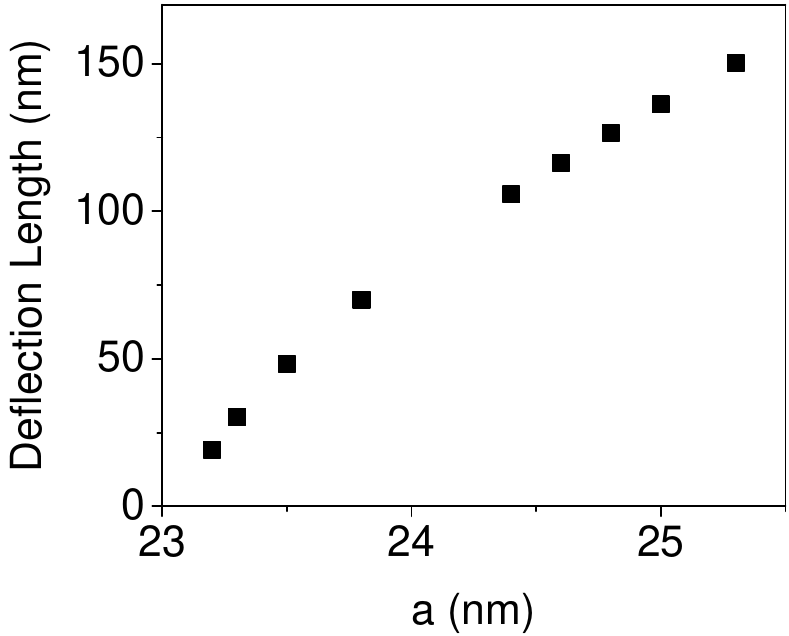}
		\caption{}
		\label{fig:Deflection}
	\end{subfigure}
	\begin{subfigure}[b]{0.55\linewidth}
			
			\includegraphics[width=\linewidth]{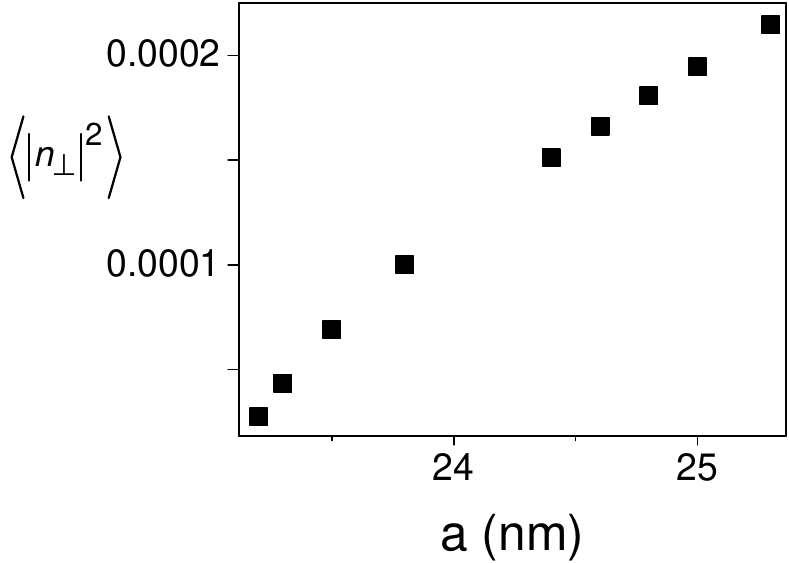}
			\caption{}
			\label{fig:fluctuations}
	\end{subfigure}
	\rule{\textwidth}{1pt}
	\caption[ElasticParameters]{Elastic constants of the hexagonal phase as a function of the lattice spacing $a$. 
	(\subref{fig:Modulus}). The bulk modulus, calculated based on the osmotic stress data (solid curve) or based on the Monte-Carlo simulation results (solid symbols).
	(\subref{fig:Deflection}). The deflection length (the average displacement between successive collisions along the chain within the confined lattice). 
	(\subref{fig:fluctuations}). The fluctuations in the nematic director. 
	}
	\label{fig:ElasticParameters}
\end{figure}

The bulk modulus:
\begin{equation}
B\equiv  -V\frac{d\Pi}{dV}
\label{bulk}
\end{equation}
can be computed from the theoretical $\Pi\left(d\right)$ (Eq. \ref{Gp}). The compressed volume, $V$, is taken to be the volume of solution per chain inside the hexagonal lattice \cite{danino2009osmotically} and is given by: $V/L=\frac{\sqrt{3}a^2}{2}-\pi\frac{D^2}{4}\equiv A_{\text{c}}$, where $L$ is the chain length and $A_{\text{c}}$ is the compressed area per chain. Assuming that $L$ remained unchanged under the osmotic pressures applied in our experiment, $dV=L\cdot dA_{\text{c}}$, hence $B=-A_{\text{c}}\frac{d\Pi}{dA_{\text{c}}}$ and is independent of $L$.  Using the above formula we can determine how the calculated bulk modulus, deflection length, and mean fluctuations in the nematic director vary with the lattice spacing $a$ (Figure \ref{fig:ElasticParameters}). The bulk modulus decreases with $a$, whereas the deflection length and the  mean fluctuations in the nematic director increase with $a$.
  
The bulk modulus can also be estimated from $\kappa$ by scaling analysis. To obtain units of pressure, $\kappa$ should be divided by a length scale. The relevant length scale in this case is the deflection length, $\lambda_{\text{def}}$, which is the length scale over which the displacement of the filament are kept within the tube around the filaments which is in the original lattice site. Hence we get that
\begin{equation}
B \approx \frac{\kappa}{\lambda_{\text{def}}}
\label{bulk2}
\end{equation}
\noindent Figure \ref{fig:ElasticParameters} confirms that the bulk modulus, which was estimated from the MC simulations by scaling analysis (Eq. \ref{bulk2}), yields bulk moduli that are of the same order of magnitude as those obtained from the osmotic stress data (Eq. \ref{bulk}). 

\subsection*{Scattering from aligned flagella samples} 
\begin{figure}[h]
\centerline{\includegraphics[width=3.25in]{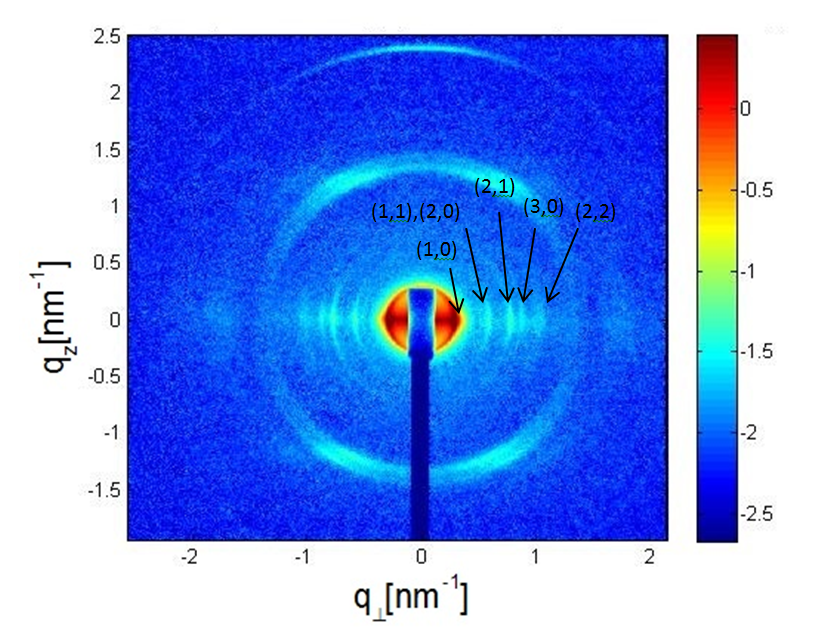}}
\caption{Two-dimensional X-ray scattering image from a partially aligned bundles of straight filaments. The filaments were aligned along their long, $z$, axis. Using Cartesian coordinates in reciprocal space, the scattering vector $\vec{q}$ has three components, $q_x, q_y$, and $q_z$. On a $2D$ detector, we can observe the $q_z$ component and the $q_\perp$ component, which is the projection of the scattering vector on the $\left(q_x, q_y\right)$ plan, and is given by $q_\perp=\sqrt{q^2_x+q^2_y}$.}
\label{2d}
\end{figure}

In few samples we obtained  high-resolution measurements of the scattering from partially aligned bundles of SJW1660 filaments. These data were taken at the ID02 beamline, ESRF, Grenoble. The $2D$ scattering data showed local bundle alignment, owing to the flow of the high bundle concentration in the narrow (ca. $\unit[2]{mm}$) quartz flow-cell capillary (Fig. \ref{2d}). The $2D$ structure-factor peaks, associated with lateral packing, were located along the perpendicular axis, $q_{\perp}$. The peak at $\unit[1.4]{nm^{-1}}$, attributed to the two turn helical pitch, was along a diagonal line situated between the vertical $\left(q_z\right)$ and horizontal $\left(q_\perp\right)$ axes, and the peak at $\unit[2.4]{nm^{-1}}$ that shows the helical axial rise, was along the axial, $q_z$, axis, as expected.

To calculate the structure-factor, $S\left(q_\perp, q_z\right)$, both the density, $\rho\left(\vec{r}\right)$ and the local displacement field $u\left(\vec{r}\right)$ should be evaluated. $u\left(\vec{r}\right)$, was calculated from the elastic free-energy for fluctuations in the hexagonal phase  \cite{Selinger1991}. $\langle\left|\vec{u}\right|^2\rangle^{\nicefrac{1}{2}}$ is directly related to the stability of the hexagonal bundle (in other words, the lattice is unstable when $\langle\left|\vec{u}\right|^2\rangle^{\nicefrac{1}{2}}$ diverge). 

\begin{figure}[h]
\centerline{\includegraphics[width=3.25in]{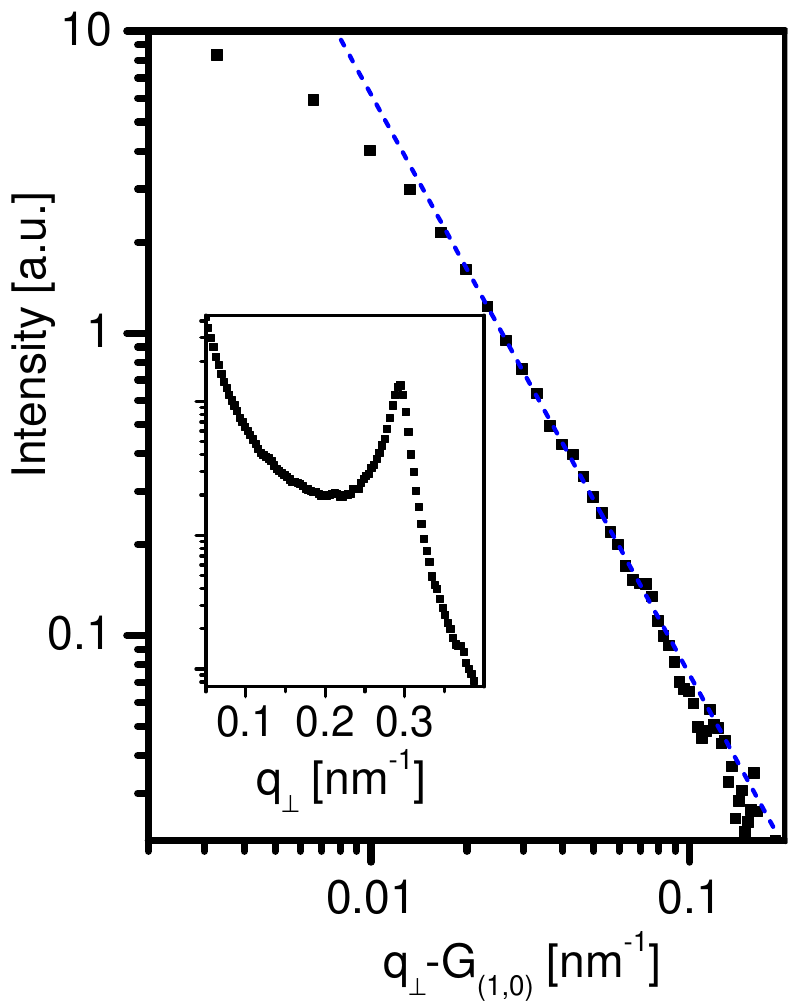}}
\caption{A log-log plot of the (1,0) peak tail (solid symbols) along $q_{\perp}$, showing a linear tail. $G_{1,0}$ is the peak center. The broken line is a linear fit with a slope of  $-2.0\pm 0.1$, as theoretically predicted \cite{Selinger1991}. The inset shows the shape of the peak on a log-linear scale.}
\label{ESRFp}
\end{figure}

\begin{figure}[h]
\centerline{\includegraphics[width=3.25in]{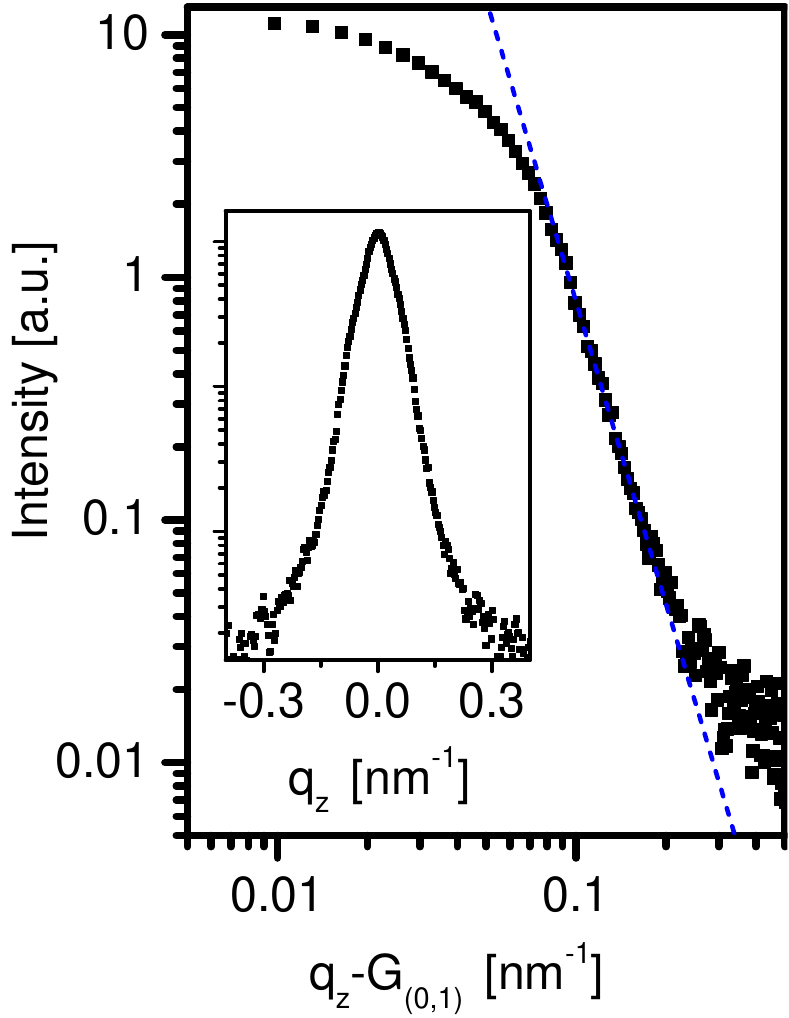}}
\caption{A log-log plot of the (0,1) peak tail (solid symbols) along $q_z$, averaged over 5 pixels in $q_\perp$, showing a linear tail. The center of the peak is at $G_{0,1}$. The broken line is a linear fit with a slope of  $-4.0\pm 0.1$, as theoretically predicted \cite{Selinger1991}. The inset shows the shape of the peak on a log-linear scale.}
\label{ESRFz}
\end{figure}

By alignment of the filaments in the flow-cell capillary we obtained two-dimensional scattering data from the hexagonal flagellar bundles that we could comparable with the theoretical structure-factor calculated for semi-flexible long chains \cite{Selinger1991}. The model, which takes into account the elastic free energy of undulations in a hexagonal phase of chains,  predicts that the line-shape of the hexagonal structure-factor peaks should decay as power laws in the tails of the peaks. In particular, for the $\left(1,0\right)$ and $\left(0,1\right)$ peaks, the data contained enough points to confirm the predicted power lows. Figure \ref{ESRFp} presents a $\log-\log$ plot of the $\left(1,0\right)$ peak along the perpendicular direction, $q_{\perp}$, with a linear fit, using a slope of $-2.0\pm 0.1$, confirming the predicated \cite{Selinger1991} structure factor line-shape of:  $SF(q_\perp-G_{\left(1,0\right)},q_z = 0)\propto q_\perp^{-2}$, where $G_{\left(1,0\right)}$ is the $\left(1,0\right)$ peak center. Figure \ref{ESRFz} presents a $\log-\log$ plot of the $\left(0,1\right)$ peak along the axial, $q_z$, axis using a linear fit with a slope of $-4.0\pm 0.1$, confirming the predicted structure-factor line-shape of: $SF(q_\perp=0, q_z - G_{\left(0,1\right)})\propto q_z^{-4}$, where $G_{\left(0,1\right)}$ is the center of the $\left(0,1\right)$ peak.  Similar structure-factor line-shapes can be expected for other filament bundles, including microtubule or neurofilament bundles \cite{safinya2013liquid}.

\section*{CONCLUSIONS}
We have used solution SAXS to determine high-resolution structure of the L-type straight flagellar filament strain SJW 1660. Using the atomic model of the flagellin subunit we calculated the scattering curve from the helical lattice of the entire filament and compared with our scattering data. We found that the helix diameter was $23.1\:\textrm{nm}$, it had a two turn pitch of $5.2\:\textrm{nm}$, and $10.96$ flagellin subunits per two turn pitch. Under osmotic stress the filaments formed $2D$ hexagonal bundles. To fit the solution X-ray scattering curves of $2D$ hexagonal bundles, Monte-Carlo simulations were used to account for thermal fluctuation effects and the interactions between filaments in the bundles, assuming harmonic pairwise potentials between neighboring filaments with an elastic constant, $\kappa$ of $\unit[1.2]{mN \cdot m^{-1}}$. 
We determined the distance between the semiflexible flagellar filaments in the bundles, as a function of osmotic stress.
We could fit the resulting pressure-distance curve to the equation-of-state of hexagonal bundles of semiflexible chains \cite{Strey1997}, from which the parameters associated with the electrostatic, hydration, and undulation interactions, were determined. The undulation energy was associated with a bending stiffness, which corresponds to a chain persistence length of $\approx \unit[700]{\mu m}$. We then computed the bundle bulk-modulus, the deflection length of the filaments within the bundle, and the mean fluctuations in the nematic director and the variation of these parameters with the hexagonal lattice spacing. Using scaling arguments we confirmed that the bundle bulk-modulus obtained from the MC simulations is in agreement with the bulk-modulus obtained from the osmotic stress data. Furthermore, the tails of the bundle structure-factor peak line-shapes followed the theoretically predicted \cite{Selinger1991} power law behavior (for semi-flexible chains) with exponents of $-2$ and $-4$ in the perpendicular and axial directions, respectively. 

\section*{SUPPLEMENTARY MATERIAL}

\ack{An online supplement to this article can be found by visiting BJ Online at http://www.biophysj.org.
}\vspace*{-3pt}
\label{Supporting Information}
%\lhead{Supporting Information}

\subsection{Solvent subtraction}
Figure \ref{SoventSubtractedFF} shows that using Eq. \ref {bsmonomer} to subtract the displaced volume of the solvent did not improve the fit to our data. Hence Eq. \ref{monomerFF} was used in subsequent models.

\subsection{Varying form-factor parameters}
\begin{figure}[h]
\centerline{\includegraphics[width=3.25in]{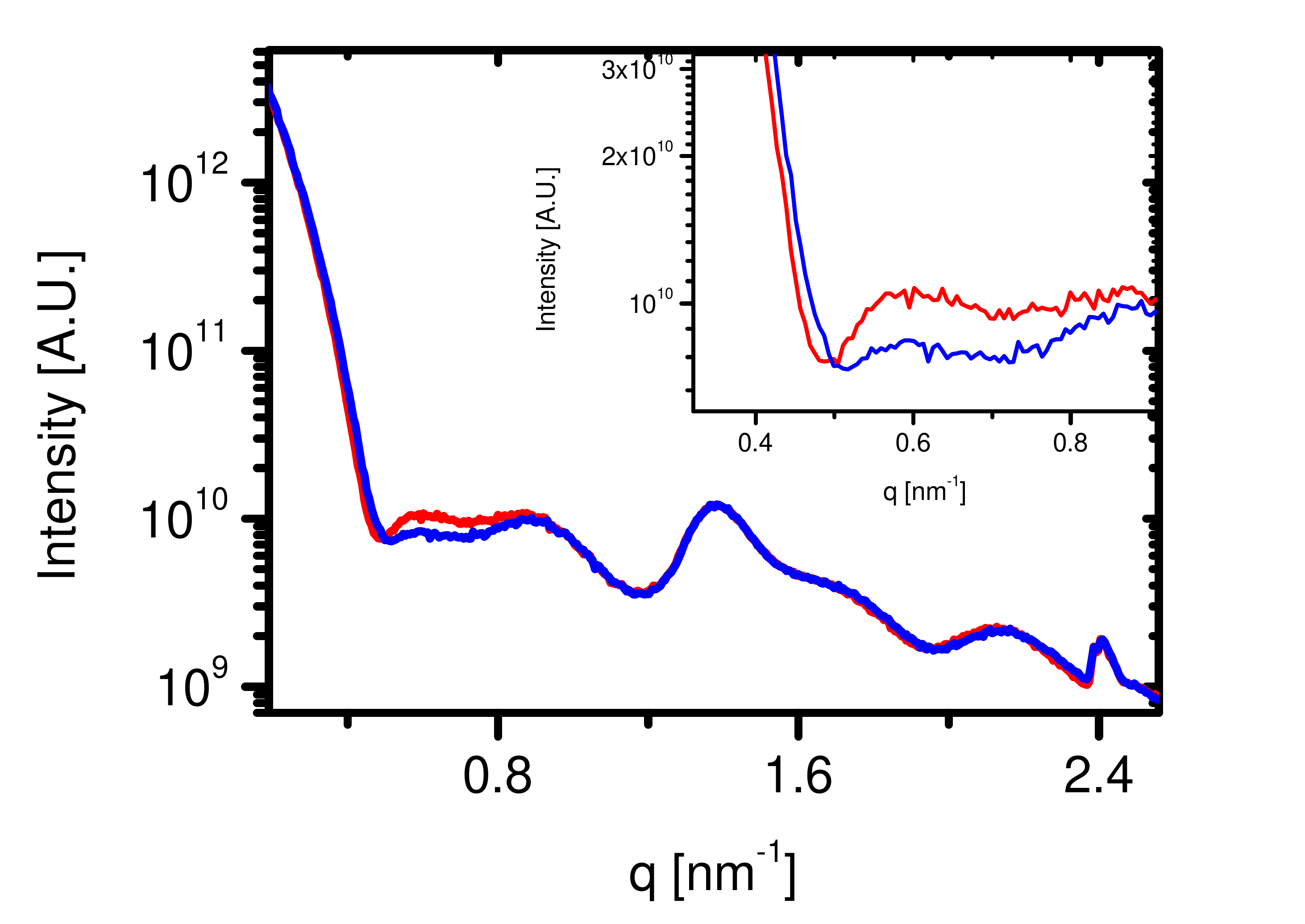}}
\caption{A comparison between two form-factor models with a $\unit[0.1]{nm}$ difference in the helix diameter. The main differences between the two models are the location of the first minimum and the amplitude of the following local maximum. The inset, shows these features on an expanded scale.}
\label{FFradius}
\end{figure}

\begin{figure}[h]
\centerline{\includegraphics[width=3.25in]{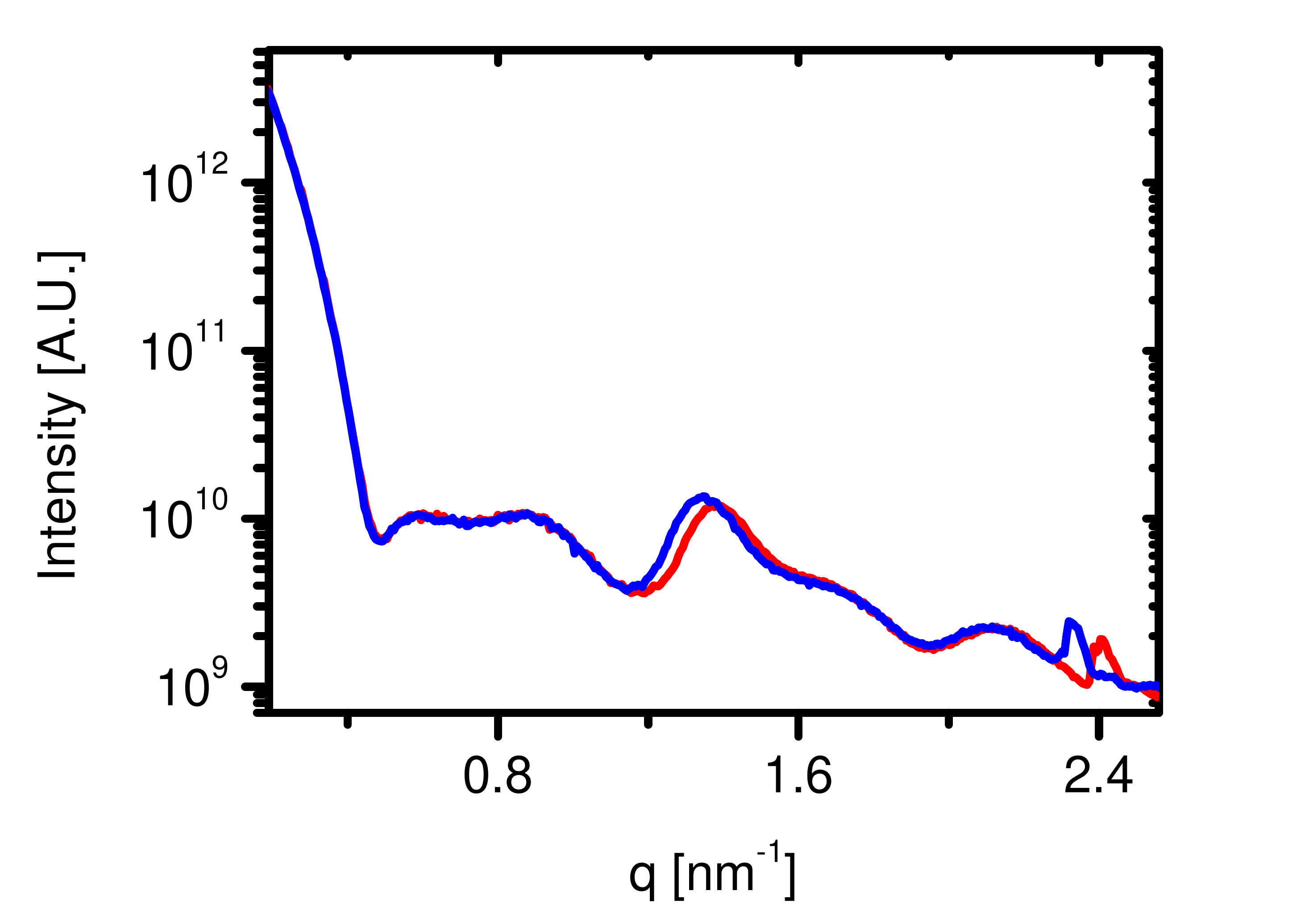}}
\caption{A comparison between two form-factor models with a $\unit[0.2]{nm}$ difference in the size of the two turn pitch. The main difference between the two models is the shift of the two layer-line peaks at $q\simeq \unit[1.4]{nm^{-1}}$ and $q\simeq \unit[2.4]{nm^{-1}}$.}
\label{FFpitch}
\end{figure}

\begin{figure}[h]
\centerline{\includegraphics[width=3.25in]{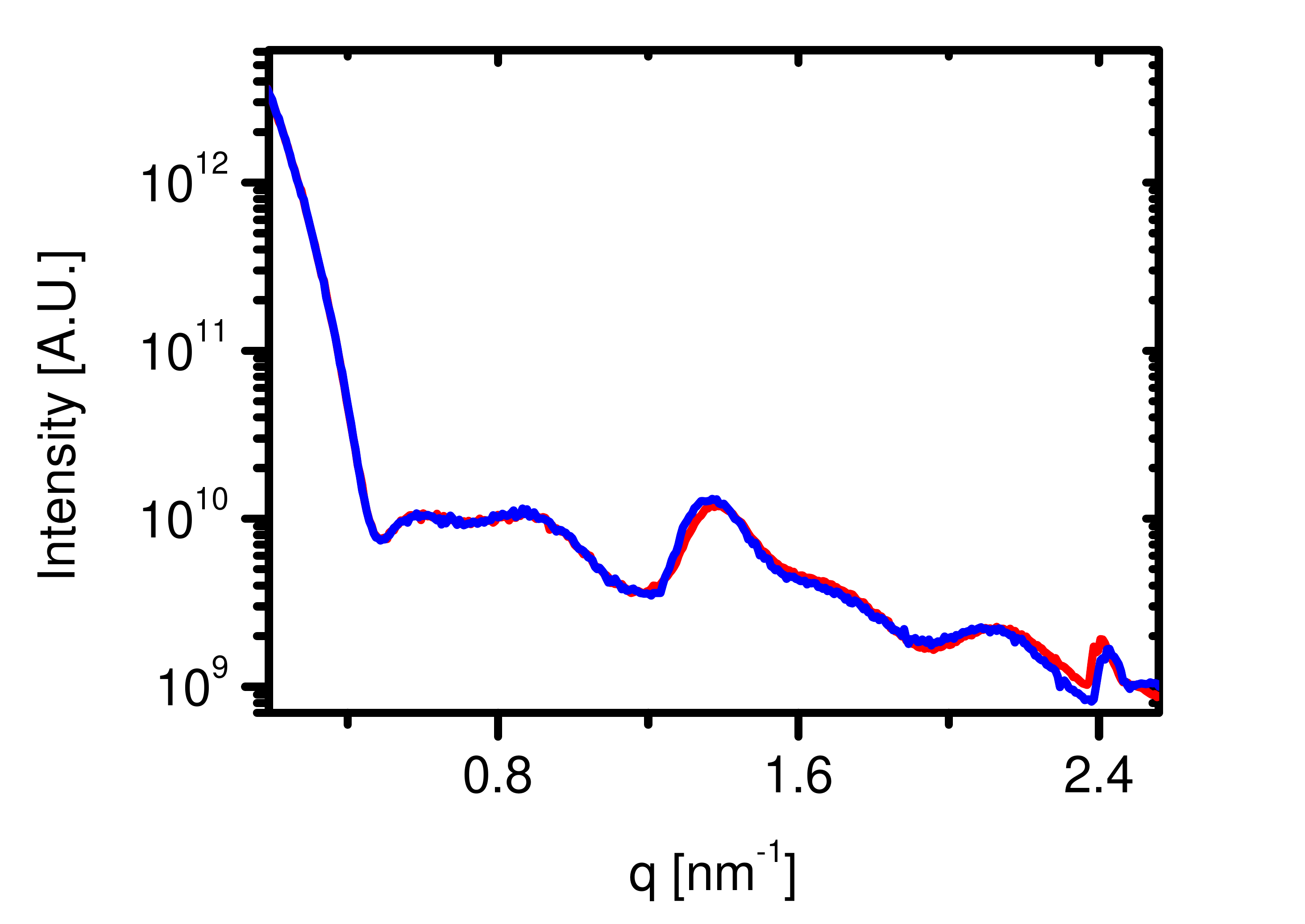}}
\caption{A comparison between two form-factor models with a $1^{\circ}$ difference in the angle of the filament tilt. The  tilt angle mainly changes the separation between the two layer-line peaks at $q\simeq \unit[1.4]{nm^{-1}}$ and $q\simeq \unit[2.4]{nm^{-1}}$.}
\label{FFtilt}
\end{figure}

The high resolution flagella model (Figure \ref{FF1660}) is supported by the following considerations. The helical character of the structure is supported by looking at the aligned sample shown in Figure \ref{2d}. In this Figure, the form factor features are governed  by the helical shape of the filament at $q>\unit[1]{nm^{-1}}$, indicating that the subunit packing must be considered in the form-factor model. 
Figures \ref{FFradius}-\ref{FFtilt} show the high sensitivity of the calculated form-factor model to small changes in the helix diameter (Figure \ref{FFradius}), the helical pitch (Figure \ref{FFpitch}), and the filament tilt angle (Figure \ref{FFtilt}).

Figures \ref{FFradius}-\ref{FFtilt} clearly show that each of these parameters predominantly affects different features in the form-factor, allowing them to be optimized (or fit) independently. This fact increase our confidence level in the model.

Figure \ref{SF0itr} shows the calculated $I(q)$ derived from the model of flagellar filaments, arranged in a hexagonal lattice with no thermal fluctuations. The lattice-sum peaks dominates this model and there are almost no form-factor features at $q>\:\unit[\sim0.5]{nm^{-1}}$.

\subsection{Instrument resolution function}
Figure \ref{SF0itr} shows that the intensity of the flagellar bundle model has sharp correlation peaks that were invisible in the SAXS results (Figure \ref{SF1660}). We attribute this observation to the resolution function of our measurement setups, defined by the monochromator, detector pixel size, beam size, sample-to-detector distance, etc. To account for these effects, each of the calculated model intensities was convoluted with a Gaussian function with a standard deviation, $\sigma=\unit[0.03]{nm^{-1}}$, which is the measured resolution of our setup. Figure \ref{SF0itr} (red curve) shows that fewer measurable peaks are expected when the resolution function is taken into account.

\subsection{Varying Structure-factor parameters}
\begin{figure}[h]
\centerline{\includegraphics[width=3.25in]{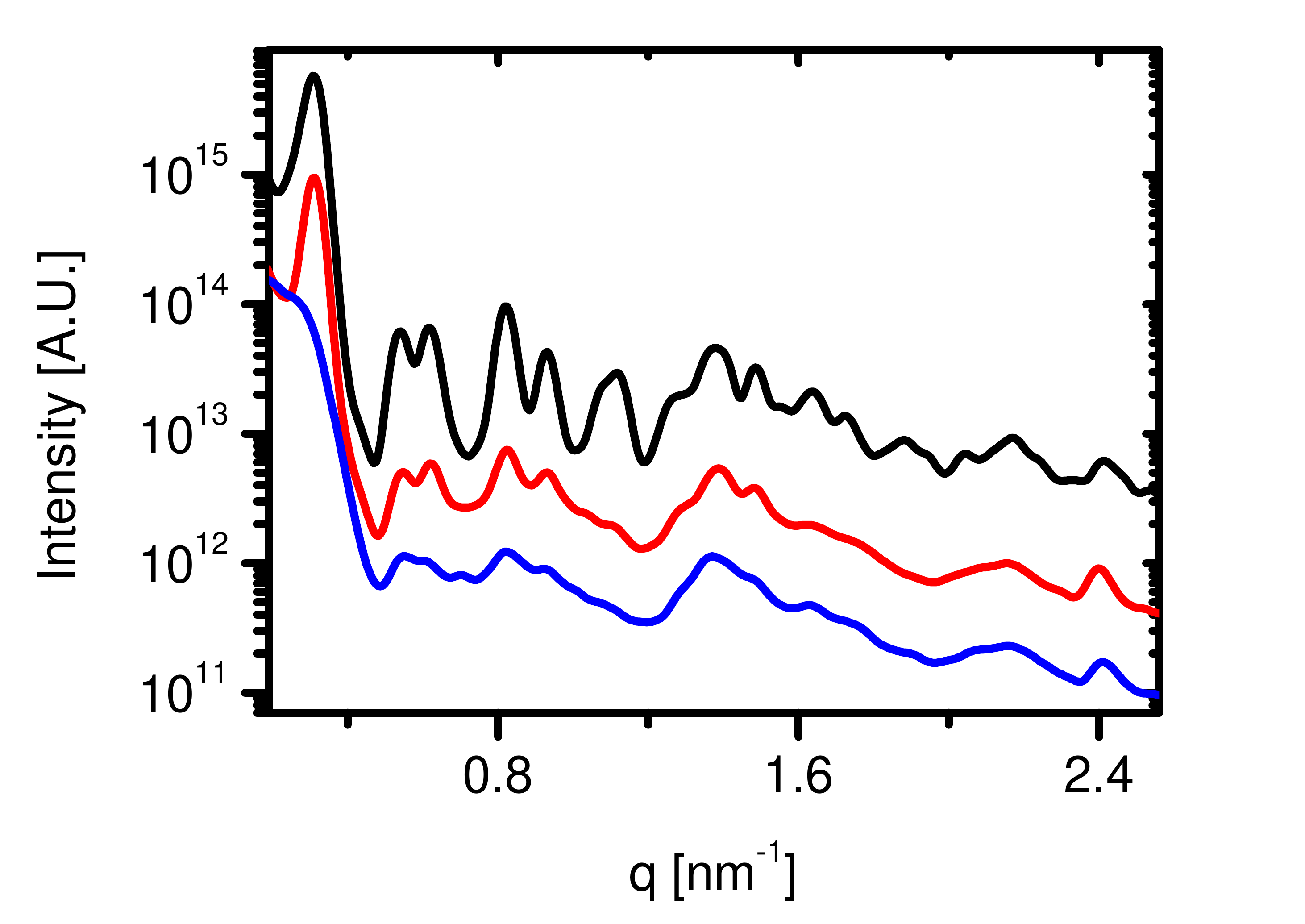}}
\caption{A comparison between the lattice sum contribution to the calculated intensity of three models with different $\kappa$ values. $\kappa=\unit[8]{mN\cdot m^{-1}}$ (black curve), $\kappa=\unit[0.8]{mN\cdot m^{-1}}$ (red curve), and  $\kappa=\unit[0.08]{mN\cdot m^{-1}}$ (blue curve).}
\label{SFcomp}
\end{figure}

The SAXS results (Figure \ref{SF1660}) show that both the structure factor peaks and the form-factor principle features can be observed. Figure \ref{SFcomp} compares between the calculated intensity models of flagellar filament hexagonal bundles with different degrees of thermal fluctuations. The extent of fluctuations was determined by the elastic constant between neighbours', $\kappa$, which determines the lattice-sum contribution to the calculated intensity. A sufficiently high $\kappa$ value, significantly limits thermal fluctuations and the intensity resembles the calculated intensity assuming no thermal fluctuations (Figure \ref{SF0itr}, blue curve). If, however, the value of $\kappa$ is too low, the hexagonal lattice is unstable and the lattice-sum peaks become unclear (Figure \ref{SFcomp}, blue curve).

The hexagonal lattice constant, $a$, affects the location of the peaks in the calculated intensity. Figure \ref{latticeComp} shows models with a small difference in the value of $a$. The Figure clearly shows that the model is very sensitive to the value of $a$, hence $a$ can be accurately determined.

Figure \ref{DomainComp} shows how the calculated intensity, $I(q)$, is affected by the size of the bundle. Here the main feature that is affected is the width of the peaks. To clearly see that, the results before and after applying the convolution with the experimental resolution function are shown.

Unlike the form-factor parameters, the lattice-sum parameters are  more dependent of each other and it is possible to attain similar fits by fine tuning $\kappa$ or the bundle size. These parameters should therefore be considered more carefully. It is, however, clear that the model provides the correct order of magnitude of these parameters. 

\subsection{Osmotic stress experiments}
\label{OsmoticStress}
Figure \ref{MorePressures} provides additional SAXS curves from flagellar bundles formed under different osmotic pressures, as in Figure \ref{SF1660}. 

\subsection{Equation of state}
\label{EofS}
The equation-of-states for a bundle of long semi-flexible chains in solution is: 
\begin{equation}
\frac{\partial G}{\partial d}(d)=\frac{\partial H_0}{\partial d}(d)
+ck_BT\kappa_s^{-\frac{1}{4}}\frac{\partial}{\partial
  d}\sqrt[4]{\frac{\partial^2 H_0}{\partial d^2}}
\label{statesS}
\end{equation}
where G is the free energy, $d = a - D$ is the spacing between filaments,
$\kappa_s$ is the bending stiffness, $H_0$ is:
\begin{equation}
H_0(d)=a_h\frac{e^{-d/\lambda_H}}{\sqrt{d/\lambda_H}}+b\frac{e^{-d/\lambda_D}}{\sqrt{d/\lambda_D}},
\label{H0S}
\end{equation}
where $\lambda_H$ and $\lambda_D$ are the hydration and electrostatic screening lengths. 
The first derivative of $H_0$ is:
\begin{equation}
\frac{\partial H_0}{\partial d}(d)=-\frac{a_h\left(\frac{2d}{\lambda_H}+1\right)\mathrm{e}^{-\frac{d}{\lambda_H}}}{2\lambda_H\left(\frac{d}{\lambda_H}\right)^\frac{3}{2}}-\frac{b\left(\frac{2d}{\lambda_D}+1\right)\mathrm{e}^{-\frac{d}{\lambda_D}}}{2\lambda_D\left(\frac{d}{\lambda_D}\right)^\frac{3}{2}}
\end{equation}
and its second derivative is:
\begin{equation}
\frac{\partial^2 H_0}{\partial d^2}(d)=\frac{a_h\left(\frac{4d^2}{\lambda_H}+4d+3\lambda_H\right)\mathrm{e}^{-\frac{d}{\lambda_H}}}{4\lambda_H^3\left(\frac{d}{\lambda_H}\right)^\frac{5}{2}}+\frac{b\left(\frac{4d^2}{\lambda_D}+4d+3\lambda_D\right)\mathrm{e}^{-\frac{d}{\lambda_D}}}{4\lambda_D^3\left(\frac{d}{\lambda_D}\right)^\frac{5}{2}}
\label{d2H0}
\end{equation}
The last term of the equation of state is then:
\begin{gather}
\frac{\partial}{\partial
  d}\sqrt[4]{\frac{\partial^2 H_0}{\partial d^2}}=-\frac{\left(\alpha+
  \beta\right)\mathrm{e}^{-\frac{d}{\lambda_H}-\frac{d}{\lambda_D}}}{\gamma}
\end{gather}

where 
\begin{equation}
\alpha=\left(\frac{d}{\lambda_H}\right)^\frac{7}{2}\left(8b\lambda_H^6d^3+12b\lambda_D\lambda_H^6d^2+18b\lambda_D^2\lambda_H^6x+15b\lambda_D^3\lambda_H^6\right)\mathrm{e}^\frac{d}{\lambda_H}
\end{equation}
\begin{equation}
\beta=\left(\frac{d}{\lambda_D}\right)^\frac{7}{2}\left(8a\lambda_D^6d^3+12a_h\lambda_D^6\lambda_Hd^2+18a_h\lambda_D^6\lambda_H^2d+15a_h\lambda_D^6\lambda_H^3\right)\mathrm{e}^\frac{d}{\lambda_D}
\end{equation}
and 
\begin{equation}
\gamma=32\lambda_D^6\lambda_H^6\left(\frac{d}{\lambda_D}\right)^\frac{7}{2}\left(\frac{d}{l}\right)^\frac{7}{2}\left(\frac{a_h\left(\frac{4d^2}{\lambda_H}+4d+3\lambda_H\right)\mathrm{e}^{-\frac{d}{\lambda_H}}}{4\lambda_H^3\left(\frac{d}{\lambda_H}\right)^\frac{5}{2}}+\frac{b\left(\frac{4d^2}{\lambda_D}+4d+3\lambda_D\right)\mathrm{e}^{-\frac{d}{\lambda_D}}}{4\lambda_D^3\left(\frac{d}{\lambda_D}\right)^\frac{5}{2}}\right)^\frac{3}{4}
\end{equation}
In a hexagonal lattice, the relation between the free energy and the osmotic pressure is:

\begin{equation}
-\frac{\partial G}{\partial d}(d)=\sqrt{3}\Pi d.
\label{Gp}
\end{equation}

\section*{ACKNOWLEDGMENTS}

\ack{We thank Daniel Harries and Daniel J. Needleman for helpful discussions. We acknowledge use of ID02 beamline at ESRF where some of our SAXS data were acquired. We thank T. Narayanan and G. Lotze for their help with these measurements. DL, AG TD and UR acknowledges financial support from the Israel Science Foundation (grant 1372/13), US-Israel binational Science Foundation (grant 2009271), Rudin, Wolfson, and Safra foundations as well as the FTA-Hybrid Nanomaterials program of the Planning and Budgeting Committee of the Israel Council of Higher Education. D.L. and T.D. thank the Nanocenter of the Hebrew University for fellowships. A.G. thanks the Institute for Drug Research at the Hebrew University for a fellowship. ZD and WS acknowledge support of National Science Foundation through grants DMR-CMMI-1068566, NSF-DMR-1609742 and NSF-MRSEC-1420382. We also acknowledge use of MRSEC Biosynthesis facility supported by grant NSF-MRSEC-1420382. ZD, WS and UR acknowledge travel support from the Bronfman foundation.}

\section*{Author Contributions}
DL, ZD, and UR designed research; DL, AG, TD, WS, ZD, and UR performed research; DL, AG, TD, and UR contributed analytic tools; DL, ZD and UR analyzed data; DL, ZD and UR wrote the manuscript.


% Here references are directly included this tex file.
% But we can generate reference list from bibliography database
% Compile and format the bibliography (bj_bibtex_template.bib BibTeX
% file must be present in the document directory)

%The source file for this document is called
%\emph{biophys\_latex\_template.tex}.  Apart from this \LaTeX\ file, you
%will also need the bibliography file, the \BibTeX\ style file, and the
%EPS and PDF figure files.

%See the bibliography file \emph{bj\_bibtex\_template.bib} for the
%literature data.  It was mostly generated from the saved
%text-formatted PubMed entries using the \emph{med2bib} program and
%edited by the \emph{tkbibtex} or directly in the \emph{emacs} editor.

%The \emph{biophysj.bst} file is a \BibTeX\ style file that contains
%information about the format required by Biophysical Journal for the
%list of references.

\bibliographystyle{biophysj}
\bibliography{bj_bibtex_template}

\begin{thebibliography}{51}
\providecommand{\url}[1]{\texttt{#1}}
\providecommand{\urlprefix}{ }

\bibitem[Berg(2000)]{berg2000}
Berg, H., 2000.
\newblock Motile behavior of bacteria.
\newblock \emph{Physics Today} 53:24 -- 29.
\newblock \urlprefix\url{http://w3.impa.br/~jair/ptoday1.html}.

\bibitem[C.R.(1975)]{Calladine1975}
C.R., C., 1975.
\newblock Construction of bacterial flagella.
\newblock \emph{Nature} 255:121 -- 124.

\bibitem[Asakura(1969)]{Asakura1970}
Asakura, S., 1969.
\newblock Polymerization of flagellin and polymorphism of flagella.
\newblock \emph{Advances in biophysics} 1:99--155.

\bibitem[Hasegawa et~al.(1998{\natexlab{a}})Hasegawa, Yamashita, and
  Namba]{Hasegawa1998}
Hasegawa, K., I.~Yamashita, and K.~Namba, 1998.
\newblock Quasi- and Nonequivalence in the Structure of Bacterial Flagellar
  Filament.
\newblock \emph{Biophysical Journal} 74:569 -- 575.
\newblock
  \urlprefix\url{http://www.sciencedirect.com/science/article/pii/S00063495987%
78154}.

\bibitem[Hyman and Trachtenberg(1991)]{Hyman1991}
Hyman, H.~C., and S.~Trachtenberg, 1991.
\newblock Point mutations that lock $Salmonella typhimurium$ flagellar
  filaments in the straight right-handed and left-handed forms and their
  relation to filament superhelicity.
\newblock \emph{Journal of molecular biology} 220:79--88.

\bibitem[Kamiya et~al.(1979)Kamiya, Asakura, Wakabayashi, and
  Namba]{Kamiya1979}
Kamiya, R., S.~Asakura, K.~Wakabayashi, and K.~Namba, 1979.
\newblock Transition of bacterial flagella from helical to straight forms with
  different subunit arrangements.
\newblock \emph{Journal of Molecular Biology} 131:725 -- 742.
\newblock
  \urlprefix\url{http://www.sciencedirect.com/science/article/pii/002228367990%
1992}.

\bibitem[Trachtenberg and DeRosier(1991)]{Trachtenberg1991}
Trachtenberg, S., and D.~J. DeRosier, 1991.
\newblock A molecular switch: Subunit rotations involved in the right-handed to
  left-handed transitions of $Salmonella typhimurium$ flagellar filaments.
\newblock \emph{Journal of molecular biology} 220:67--77.

\bibitem[Darnton and Berg(2007)]{Darnton2007}
Darnton, N.~C., and H.~C. Berg, 2007.
\newblock Force-extension measurements on bacterial flagella: triggering
  polymorphic transformations.
\newblock \emph{Biophysical journal} 92:2230--2236.

\bibitem[Srigiriraju and Powers(2005)]{Srigiriraju2005}
Srigiriraju, S.~V., and T.~R. Powers, 2005.
\newblock Continuum Model for Polymorphism of Bacterial Flagella.
\newblock \emph{Phys. Rev. Lett.} 94:248101.
\newblock
  \urlprefix\url{http://link.aps.org/doi/10.1103/PhysRevLett.94.248101}.

\bibitem[Barry et~al.(2006)Barry, Hensel, Dogic, Shribak, and
  Oldenbourg]{Barry2006}
Barry, E., Z.~Hensel, Z.~Dogic, M.~Shribak, and R.~Oldenbourg, 2006.
\newblock Entropy-Driven Formation of a Chiral Liquid-Crystalline Phase of
  Helical Filaments.
\newblock \emph{Phys. Rev. Lett.} 96:018305.
\newblock
  \urlprefix\url{http://link.aps.org/doi/10.1103/PhysRevLett.96.018305}.

\bibitem[Trachtenberg and Hammel(1992)]{Trachtenberg1992}
Trachtenberg, S., and I.~Hammel, 1992.
\newblock The rigidity of bacterial flagellar filaments and its relation to
  filament polymorphism.
\newblock \emph{Journal of structural biology} 109:18--27.

\bibitem[Hoshikawa and Kamiya(1985)]{Hoshikawa1985}
Hoshikawa, H., and R.~Kamiya, 1985.
\newblock Elastic properties of bacterial flagellar filaments: II.
  Determination of the modulus of rigidity.
\newblock \emph{Biophysical chemistry} 22:159--166.

\bibitem[Selinger and Bruinsma(1991)]{Selinger1991}
Selinger, J.~V., and R.~F. Bruinsma, 1991.
\newblock Hexagonal and nematic phases of chains. I. Correlation functions.
\newblock \emph{Phys. Rev. A} 43:2910--2921.
\newblock \urlprefix\url{http://link.aps.org/doi/10.1103/PhysRevA.43.2910}.

\bibitem[Strey et~al.(1997)Strey, Parsegian, and Podgornik]{Strey1997}
Strey, H.~H., V.~A. Parsegian, and R.~Podgornik, 1997.
\newblock Equation of State for DNA Liquid Crystals: Fluctuation Enhanced
  Electrostatic Double Layer Repulsion.
\newblock \emph{Phys. Rev. Lett.} 78:895--898.
\newblock \urlprefix\url{http://link.aps.org/doi/10.1103/PhysRevLett.78.895}.

\bibitem[Maki-Yonekura et~al.(2010)Maki-Yonekura, Yonekura, and
  Namba]{maki2010conformational}
Maki-Yonekura, S., K.~Yonekura, and K.~Namba, 2010.
\newblock Conformational change of flagellin for polymorphic supercoiling of
  the flagellar filament.
\newblock \emph{Nature structural \& molecular biology} 17:417--422.

\bibitem[Steiner et~al.(2012)Steiner, Szekely, Szekely, Dvir, Asor, Yuval-Naeh,
  Keren, Kesselman, Danino, Resh, Ginsburg, Guralnik, Feldblum, Tamburu, Peres,
  and Raviv]{Steiner2012}
Steiner, A., P.~Szekely, O.~Szekely, T.~Dvir, R.~Asor, N.~Yuval-Naeh, N.~Keren,
  E.~Kesselman, D.~Danino, R.~Resh, A.~Ginsburg, V.~Guralnik, E.~Feldblum,
  C.~Tamburu, M.~Peres, and U.~Raviv, 2012.
\newblock Entropic Attraction Condenses Like-Charged Interfaces Composed of
  Self-Assembled Molecules.
\newblock \emph{Langmuir} 28:2604--2613.
\newblock \urlprefix\url{http://pubs.acs.org/doi/abs/10.1021/la203540p}.

\bibitem[Needleman et~al.(2004)Needleman, Ojeda-Lopez, Raviv, Ewert, Jones,
  Miller, Wilson, and Safinya]{needleman2004synchrotron}
Needleman, D.~J., M.~A. Ojeda-Lopez, U.~Raviv, K.~Ewert, J.~B. Jones, H.~P.
  Miller, L.~Wilson, and C.~R. Safinya, 2004.
\newblock Synchrotron X-ray diffraction study of microtubules buckling and
  bundling under osmotic stress: a probe of interprotofilament interactions.
\newblock \emph{Physical review letters} 93:198104.

\bibitem[Needleman et~al.(2005)Needleman, Ojeda-Lopez, Raviv, Ewert, Miller,
  Wilson, and Safinya]{needleman2005radial}
Needleman, D.~J., M.~A. Ojeda-Lopez, U.~Raviv, K.~Ewert, H.~P. Miller,
  L.~Wilson, and C.~R. Safinya, 2005.
\newblock Radial compression of microtubules and the mechanism of action of
  taxol and associated proteins.
\newblock \emph{Biophysical journal} 89:3410--3423.

\bibitem[Szekely et~al.(2012)Szekely, Asor, Dvir, Szekely, and
  Raviv]{Szekely2012}
Szekely, P., R.~Asor, T.~Dvir, O.~Szekely, and U.~Raviv, 2012.
\newblock Effect of Temperature on the Interactions between Dipolar Membranes.
\newblock \emph{The Journal of Physical Chemistry B} 116:3519--3524.
\newblock \urlprefix\url{http://pubs.acs.org/doi/abs/10.1021/jp209157y}.

\bibitem[Cohen et~al.(2009)Cohen, Podgornik, Hansen, and Parsegian]{Cohen2009}
Cohen, J.~A., R.~Podgornik, P.~L. Hansen, and V.~A. Parsegian, 2009.
\newblock A Phenomenological One-Parameter Equation of State for Osmotic
  Pressures of PEG and Other Neutral Flexible Polymers in Good Solvents.
\newblock \emph{The Journal of Physical Chemistry B} 113:3709--3714.
\newblock \urlprefix\url{http://pubs.acs.org/doi/abs/10.1021/jp806893a}, pMID:
  19265418.

\bibitem[Nadler et~al.(2011)Nadler, Steiner, Dvir, Szekely, Szekely, Ginsburg,
  Asor, Resh, Tamburu, Peres, and Raviv]{Nadler2011}
Nadler, M., A.~Steiner, T.~Dvir, O.~Szekely, P.~Szekely, A.~Ginsburg, R.~Asor,
  R.~Resh, C.~Tamburu, M.~Peres, and U.~Raviv, 2011.
\newblock Following the structural changes during zinc-induced crystallization
  of charged membranes using time-resolved solution X-ray scattering.
\newblock \emph{Soft Matter} 7:1512--1523.
\newblock \urlprefix\url{http://dx.doi.org/10.1039/C0SM00824A}.

\bibitem[Ben-Nun et~al.(2010)Ben-Nun, Ginsburg, Sz{\'{e}}kely, and
  Raviv]{Ben-Nun2010}
Ben-Nun, T., A.~Ginsburg, P.~Sz{\'{e}}kely, and U.~Raviv, 2010.
\newblock {{\it X+}: a comprehensive computationally accelerated structure
  analysis tool for solution X-ray scattering from supramolecular
  self-assemblies}.
\newblock \emph{Journal of Applied Crystallography} 43:1522--1531.
\newblock \urlprefix\url{http://dx.doi.org/10.1107/S0021889810032772}.

\bibitem[Székely et~al.(2010)Székely, Ginsburg, Ben-Nun, and
  Raviv]{Székely2010}
Székely, P., A.~Ginsburg, T.~Ben-Nun, and U.~Raviv, 2010.
\newblock Solution X-ray Scattering Form Factors of Supramolecular
  Self-Assembled Structures.
\newblock \emph{Langmuir} 26:13110--13129.
\newblock \urlprefix\url{http://pubs.acs.org/doi/abs/10.1021/la101433t}.

\bibitem[Ben-Nun et~al.(2016{\natexlab{a}})Ben-Nun, Barak, and
  Raviv]{BenNun2016132}
Ben-Nun, T., A.~Barak, and U.~Raviv, 2016.
\newblock Spline-based parallel nonlinear optimization of function sequences.
\newblock \emph{Journal of Parallel and Distributed Computing} 93–94:132 --
  145.
\newblock
  \urlprefix\url{http://www.sciencedirect.com/science/article/pii/S07437315163%
0017X}.

\bibitem[Ben-Nun et~al.(2016{\natexlab{b}})Ben-Nun, Asor, Ginsburg, and
  Raviv]{IJCH:IJCH201500037}
Ben-Nun, T., R.~Asor, A.~Ginsburg, and U.~Raviv, 2016.
\newblock Solution X-ray Scattering Form-Factors with Arbitrary Electron
  Density Profiles and Polydispersity Distributions.
\newblock \emph{Israel Journal of Chemistry} 56:622--628.
\newblock \urlprefix\url{http://dx.doi.org/10.1002/ijch.201500037}.

\bibitem[Ginsburg et~al.(2016)Ginsburg, Ben-Nun, Asor, Shemesh, Ringel, and
  Raviv]{doi:10.1021/acs.jcim.6b00159}
Ginsburg, A., T.~Ben-Nun, R.~Asor, A.~Shemesh, I.~Ringel, and U.~Raviv, 2016.
\newblock Reciprocal Grids: A Hierarchical Algorithm for Computing Solution
  X-ray Scattering Curves from Supramolecular Complexes at High Resolution.
\newblock \emph{Journal of Chemical Information and Modeling} 56:1518--1527.
\newblock \urlprefix\url{http://dx.doi.org/10.1021/acs.jcim.6b00159}, pMID:
  27410762.

\bibitem[Mimori-Kiyosue et~al.(1996)Mimori-Kiyosue, Vonderviszt, Yamashita,
  Fujiyoshi, and Namba]{Mimori-Kiyosue1996}
Mimori-Kiyosue, Y., F.~Vonderviszt, I.~Yamashita, Y.~Fujiyoshi, and K.~Namba,
  1996.
\newblock Direct interaction of flagellin termini essential for polymorphic
  ability of flagellar filament.
\newblock \emph{Proceedings of the National Academy of Sciences}
  93:15108--15113.
\newblock \urlprefix\url{http://www.pnas.org/content/93/26/15108.abstract}.

\bibitem[Yonekura et~al.(2003)Yonekura, Maki-Yonekura, and Namba]{Yonekura2003}
Yonekura, K., S.~Maki-Yonekura, and K.~Namba, 2003.
\newblock Complete atomic model of the bacterial flagellar filament by electron
  cryomicroscopy.
\newblock \emph{Nature} 424:643--650.

\bibitem[Hamilton(1974)]{hamilton1974international}
Hamilton, W., 1974.
\newblock International Tables for X-ray Crystallography, vol. IV.
\newblock \emph{Birmingham: Kynoch Press.(Present distributor Kluwer Academic
  Publishers, Dordrecht.)} 273--284.

\bibitem[Marsh and Slagle(1983)]{marsh1983corrections}
Marsh, R., and K.~Slagle, 1983.
\newblock Corrections to Table 2.2 B of Volume IV of International Tables for
  X-ray Crystallography.
\newblock \emph{Acta Crystallographica Section A: Foundations of
  Crystallography} 39:173--173.

\bibitem[Svergun et~al.(1995)Svergun, Barberato, and Koch]{svergun1995crysol}
Svergun, D., C.~Barberato, and M.~Koch, 1995.
\newblock CRYSOL-a program to evaluate X-ray solution scattering of biological
  macromolecules from atomic coordinates.
\newblock \emph{Journal of Applied Crystallography} 28:768--773.

\bibitem[Koutsioubas and P{\'{e}}rez(2013)]{Koutsioubas:kk5146}
Koutsioubas, A., and J.~P{\'{e}}rez, 2013.
\newblock {Incorporation of a hydration layer in the `dummy atom' {\it ab
  initio} structural modelling of biological macromolecules}.
\newblock \emph{Journal of Applied Crystallography} 46:1884--1888.
\newblock \urlprefix\url{http://dx.doi.org/10.1107/S0021889813025387}.

\bibitem[Schneidman-Duhovny et~al.(2013)Schneidman-Duhovny, Hammel, Tainer, and
  Sali]{SchneidmanDuhovny2013962}
Schneidman-Duhovny, D., M.~Hammel, J.~Tainer, and A.~Sali, 2013.
\newblock Accurate \{SAXS\} Profile Computation and its Assessment by Contrast
  Variation Experiments.
\newblock \emph{Biophysical Journal} 105:962 -- 974.
\newblock
  \urlprefix\url{http://www.sciencedirect.com/science/article/pii/S00063495130%
08059}.

\bibitem[Fraser et~al.(1978)Fraser, MacRae, and Suzuki]{fraser1978improved}
Fraser, R., T.~MacRae, and E.~Suzuki, 1978.
\newblock An improved method for calculating the contribution of solvent to the
  X-ray diffraction pattern of biological molecules.
\newblock \emph{Journal of Applied Crystallography} 11:693--694.

\bibitem[Slater(1964)]{slater1964atomic}
Slater, J.~C., 1964.
\newblock Atomic radii in crystals.
\newblock \emph{The Journal of Chemical Physics} 41:3199--3204.

\bibitem[Roberson and Schwertassek(1988)]{roberson1988dynamics}
Roberson, R.~E., and R.~Schwertassek, 1988.
\newblock Dynamics of multibody systems, volume~18.
\newblock Springer-Verlag Berlin.

\bibitem[Ben-Shaul(2013)]{ben2013entropy}
Ben-Shaul, A., 2013.
\newblock Entropy, energy, and bending of DNA in viral capsids.
\newblock \emph{Biophysical journal} 104:L15--L17.

\bibitem[Hasegawa et~al.(1998{\natexlab{b}})Hasegawa, Suzuki, Vonderviszt,
  Mimori-Kiyosue, Namba, et~al.]{hasegawa1998structure}
Hasegawa, K., H.~Suzuki, F.~Vonderviszt, Y.~Mimori-Kiyosue, K.~Namba, et~al.,
  1998.
\newblock Structure and switching of bacterial flagellar filaments studied by
  X-ray fiber diffraction.
\newblock \emph{Nature Structural \& Molecular Biology} 5:125--132.

\bibitem[Onsager(1949)]{onsager1949effects}
Onsager, L., 1949.
\newblock The effects of shape on the interaction of colloidal particles.
\newblock \emph{Annals of the New York Academy of Sciences} 51:627--659.

\bibitem[Fraden et~al.(1989)Fraden, Maret, Caspar, and
  Meyer]{fraden1989isotropic}
Fraden, S., G.~Maret, D.~Caspar, and R.~B. Meyer, 1989.
\newblock Isotropic-nematic phase transition and angular correlations in
  isotropic suspensions of tobacco mosaic virus.
\newblock \emph{Physical review letters} 63:2068.

\bibitem[Wensink and Vroege(2003)]{wensink2003isotropic}
Wensink, H., and G.~Vroege, 2003.
\newblock Isotropic--nematic phase behavior of length-polydisperse hard rods.
\newblock \emph{The Journal of chemical physics} 119:6868--6882.

\bibitem[Stroobants et~al.(1986)Stroobants, Lekkerkerker, and
  Odijk]{stroobants1986effect}
Stroobants, A., H.~Lekkerkerker, and T.~Odijk, 1986.
\newblock Effect of electrostatic interaction on the liquid crystal phase
  transition in solutions of rodlike polyelectrolytes.
\newblock \emph{Macromolecules} 19:2232--2238.

\bibitem[Warren(1941)]{warren1941x}
Warren, B., 1941.
\newblock X-ray diffraction in random layer lattices.
\newblock \emph{Physical Review} 59:693.

\bibitem[Leikin et~al.(1993)Leikin, Parsegian, Rau, and Rand]{Parsegian1993}
Leikin, S., V.~A. Parsegian, D.~C. Rau, and R.~P. Rand, 1993.
\newblock Hydration Forces.
\newblock \emph{Annual Review of Physical Chemistry} 44:369--395.
\newblock
  \urlprefix\url{http://www.annualreviews.org/doi/abs/10.1146/annurev.pc.44.10%
0193.002101}, pMID: 8257560.

\bibitem[Israelachvili(2011)]{Jacob2011}
Israelachvili, J.~N., 2011.
\newblock Intermolecular and surface forces: revised third edition.
\newblock Academic press.

\bibitem[Gittes et~al.(1993)Gittes, Mickey, Nettleton, and
  Howard]{gittes1993flexural}
Gittes, F., B.~Mickey, J.~Nettleton, and J.~Howard, 1993.
\newblock Flexural rigidity of microtubules and actin filaments measured from
  thermal fluctuations in shape.
\newblock \emph{The Journal of cell biology} 120:923--934.

\bibitem[Odijk(1983)]{odijk1983statistics}
Odijk, T., 1983.
\newblock The statistics and dynamics of confined or entangled stiff polymers.
\newblock \emph{Macromolecules} 16:1340--1344.

\bibitem[Odijk(1986)]{odijk1986theory}
Odijk, T., 1986.
\newblock Theory of lyotropic polymer liquid crystals.
\newblock \emph{Macromolecules} 19:2313--2329.

\bibitem[Dijkstra et~al.(1993)Dijkstra, Frenkel, and
  Lekkerkerker]{DIJKSTRA1993374}
Dijkstra, M., D.~Frenkel, and H.~N. Lekkerkerker, 1993.
\newblock Confinement free energy of semiflexible polymers.
\newblock \emph{Physica A: Statistical Mechanics and its Applications} 193:374
  -- 393.
\newblock
  \urlprefix\url{http://www.sciencedirect.com/science/article/pii/037843719390%
482J}.

\bibitem[Danino et~al.(2009)Danino, Kesselman, Saper, Petrache, and
  Harries]{danino2009osmotically}
Danino, D., E.~Kesselman, G.~Saper, H.~I. Petrache, and D.~Harries, 2009.
\newblock Osmotically induced reversible transitions in lipid-DNA mesophases.
\newblock \emph{Biophysical journal} 96:L43--L45.

\bibitem[Safinya et~al.(2013)Safinya, Deek, Beck, Jones, Leal, Ewert, and
  Li]{safinya2013liquid}
Safinya, C.~R., J.~Deek, R.~Beck, J.~B. Jones, C.~Leal, K.~K. Ewert, and Y.~Li,
  2013.
\newblock Liquid crystal assemblies in biologically inspired systems.
\newblock \emph{Liquid crystals} 40:1748--1758.

\end{thebibliography}

% Bibliography style (requires the style file biophysj.bst in the
% document directory)
%\bibliographystyle{biophysj}

% Figure legends
%%Automatically it will add the figure legends  and table legends as a list by below command

\newpage

%\listoffigures

\newpage

%\listoftables

% Figures and Tables coding should be placed where the
% first reference in the text.
% All the Figure files should be placed same working directory,
% for example (fig_1.eps and fig_1.pdf files must be present
% in the document directory)

% closing statement, nothing below matters

\end{document}